\documentclass{aa}
\usepackage{graphicx} 

\usepackage{booktabs}
\usepackage[utf8]{inputenc}
\usepackage[T1]{fontenc}
\usepackage{natbib}    
\usepackage{subcaption} 
\usepackage{xcolor}
\usepackage{multirow}

\usepackage{float} 
\begin{document}
\title{A distance-independent constraint on the axion-electron coupling from RGB stars}
\titlerunning{A distance-independent constraint on the axion-electron coupling}

\author{Thomas Levasseur \inst{1,2}, Oscar Straniero \inst{1,3}, Andrea Caputo\inst{2,3, 4}} 

\institute{INAF, Osservatorio d’Abruzzo, Via M. Maggini, 64100 Teramo, Italy\\      
    \and
Dipartimento di Fisica,  Università Sapienza, Pz. A. Moro 2, 00185 Roma, Italy\\ 
     \and
INFN, Sezione di Roma, Pz. A. Moro 2, 00185 Roma, Italy\\     
     \and
Theoretical Physics Department, CERN, 1211 Geneva 23, Switzerland
             }
\date{September 2026\hfill\mbox{CERN-TH-2026-207}}

\authorrunning{T. Levasseur, O. Straniero, A. Caputo}

\abstract
{Interest in axions and axion-like particles has resurged, driven by their role in solving the strong CP problem and their appeal as dark matter candidates. The luminosity of the tip of the red giant branch (TRGB) offers a key observable for probing these particle properties.}
{This work aims to improve existing bounds on the axion-electron coupling by adopting a differential observable that is significantly less sensitive to distance, interstellar extinction, the zero-point of bolometric corrections (BCs), and other major sources of systematic uncertainty.}
{We use the bolometric magnitude difference between the TRGB and the RGB bump (RGBB) as a distance-independent constraint, applied to three globular clusters of intermediate-to-high metallicity: NGC 104 (47 Tuc), NGC 362, and NGC 5904 (M5). Using photometric catalogs of RGB stars, we determine V- and I-band magnitudes of both features and convert them to bolometric values. After validating that our stellar models reproduce the observed RGBB luminosity, we perform a maximum likelihood analysis with Monte Carlo simulations to propagate uncertainties and derive new bounds on the coupling.}
{A combined analysis of the three clusters yields a maximum likelihood at $g_{13}=g_{ae}/10^{-13} = 0.8$ and a 95\% C.L. upper limit of $1.49$. Although slightly less stringent than recent bounds from larger multi-cluster samples, this limit is substantially more robust. Moreover, under reasonable mass-loss assumptions, $g_{13} \gtrsim 7.5$ is definitively ruled out, as it predicts the disappearance of the HB and AGB phases routinely observed in globular clusters, a limit more than an order of magnitude stronger than current direct experimental bounds such as XENONnT.}
{We demonstrate the effectiveness of this differential, distance-independent method for constraining physics beyond the Standard Model. Applying it to a wider sample of globular clusters would further refine the constraint on $g_{ae}$.}
 

\maketitle

\section{Introduction}\label{sect:intro} 

One of the long-standing puzzles of the Standard Model (SM) is the strong CP problem: while quantum chromodynamics (QCD) allows for CP violation through the $\theta$ term, experimental limits on the neutron electric dipole moment require the corresponding effective parameter to be extraordinarily small. The most compelling solution to this puzzle is the Peccei--Quinn (PQ) mechanism \citep{Peccei-Quinn1,Peccei-Quinn2}, which introduces a new anomalous global $U(1)_{\rm PQ}$ symmetry that is spontaneously broken. \citet{Weinberg} and \citet{Wilczek} subsequently realized that this symmetry breaking gives rise to a pseudo-Nambu--Goldstone boson, the axion, whose dynamics naturally relax the effective QCD vacuum angle to zero. More generally, axion-like particles (ALPs)\footnote{In the rest of this article, we use the generic term axion to refer to both QCD axions and axion-like particles.} arise in many extensions of the SM, notably in string theory \citep{Svrcek_2006}, and constitute well-motivated candidates for dark matter \citep{ABBOTT1983133,PRESKILL1983127,DINE1983137,OHare:2024nmr}.

The past decade has witnessed a remarkable expansion of the theoretical and experimental effort devoted to axions, with an increasingly diverse program of laboratory experiments, astrophysical observations, and cosmological probes exploring their parameter space~\cite{DiLuzio:2020wdo, Irastorza:2021tdu}. Astrophysical systems are particularly powerful laboratories for such feebly interacting particles, often probing couplings far beyond the reach of terrestrial experiments.

Axions can interact with SM fields through several effective couplings, which provide a variety of mechanisms for their production~\citep{raffelt_book}. In stellar environments, the most relevant interactions are typically those with photons and electrons. Depending on the underlying axion couplings and the properties of the stellar plasma, axions can be produced through the Primakoff conversion of photons in the screened electromagnetic fields of charged particles, photon--axion conversion in macroscopic electromagnetic fields, electron--ion and electron--electron bremsstrahlung, Compton-like scattering, electron--positron annihilation, and photon coalescence. These processes can lead to an additional channel of energy loss from stellar interiors and thereby modify stellar evolution. We refer to \citet{raffelt_book} and, for a recent discussion, \citet{caputoraffelt_2024} and \citet{carenza_2025PhyR} for comprehensive treatments of the relevant production mechanisms and astrophysical constraints.

For the small couplings of interest here, axions interact so weakly with stellar matter that, once produced, they typically free-stream out of the star, providing an additional channel for energy loss. This picture applies particularly well to light axions, whose masses are small compared with the characteristic thermal energies in stellar interiors. At larger masses, axion production becomes increasingly suppressed by the limited thermal energy available in the plasma, while sufficiently massive axions may also decay before escaping the star.

\begin{figure}
    \centering
      \includegraphics[width=\linewidth]{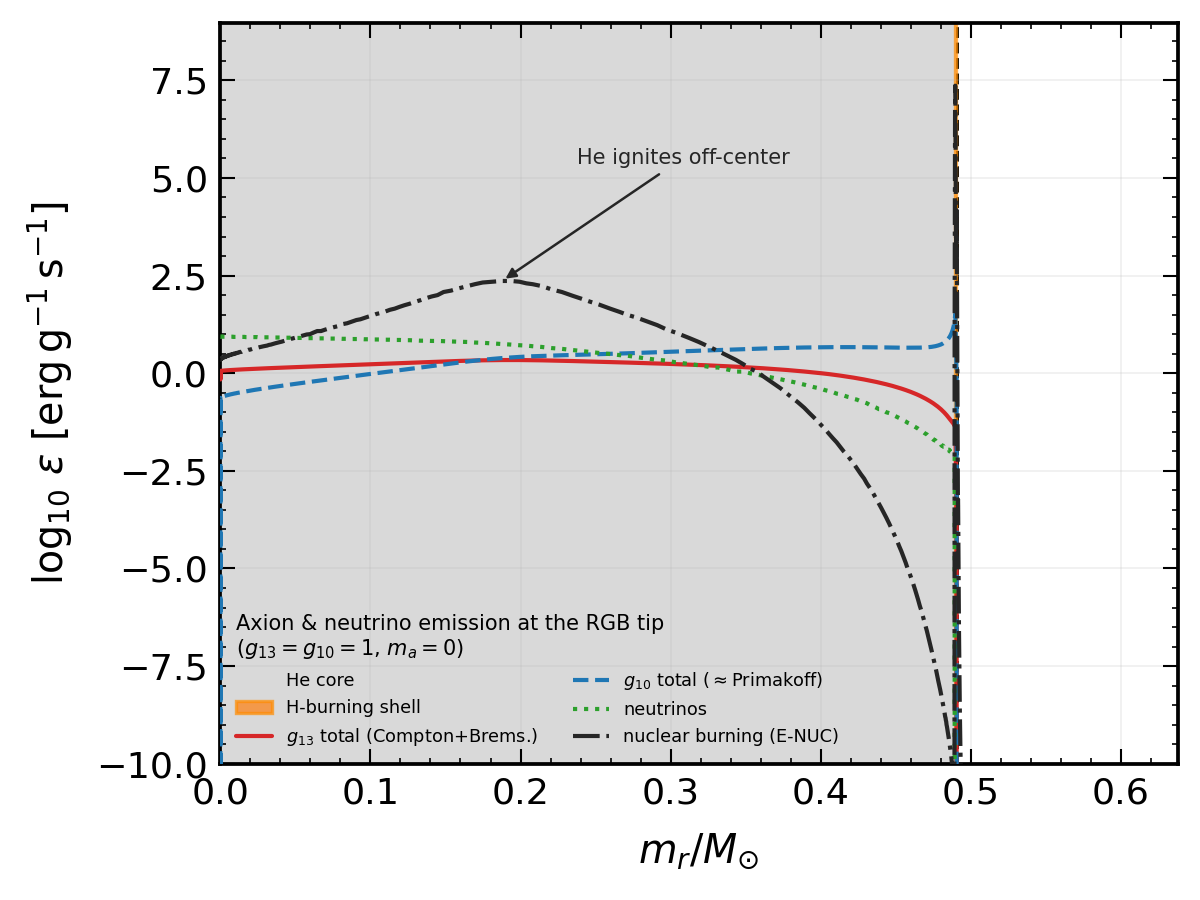} 
    \caption{Nuclear energy production and particle energy loss at the TRGB for a massless axion with $g_{13}=g_{10}=1$. The rates are computed for a $0.86M_{\odot}$ star with $Y=0.25$ and $Z=2.0\times10^{-3}$.}
    \label{fig:star_rate_profile}
\end{figure}

\begin{figure}
    \centering
    \includegraphics[width=\linewidth]{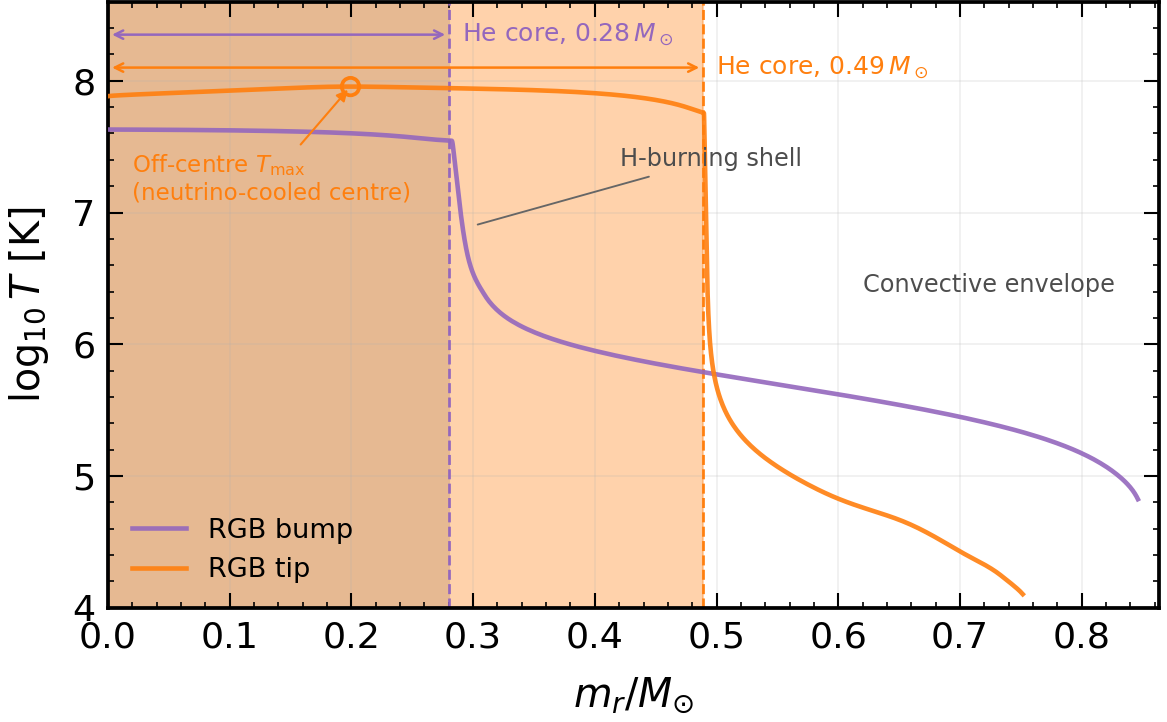}
    
    \caption{Stellar profile of $\log T$ at the TRGB and RGBB for a $0.86M_{\odot}$  star with $Y=0.25$ and $Z=2.0\times10^{-3}$.}
    \label{fig:logT_profile}
\end{figure}

\begin{figure}
    \centering
    \includegraphics[width=\linewidth]{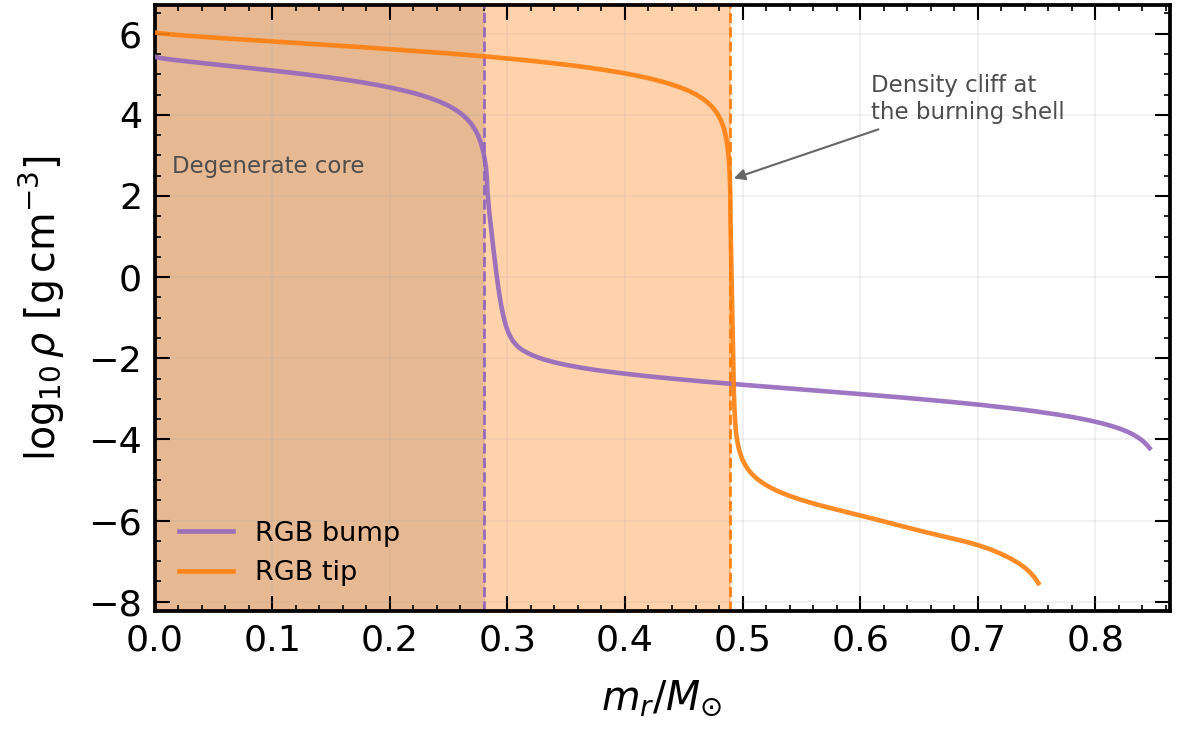}
    
    \caption{Stellar profile of $\log \rho$ at the TRGB and RGBB for a $0.86M_{\odot}$  star with $Y=0.25$ and $Z=2.0\times10^{-3}$.}
    \label{fig:log_rho_profile}
\end{figure}

In this work, we focus on the production of very light axions in red-giant-branch (RGB) stars and investigate how the associated energy loss can affect their evolution. Like neutrinos, the axion free stream from the hot core of evolved stars implies a net energy loss. Therefore, it can be accounted for by adding a term $\epsilon_{a}$ to the energy conservation equation:
\begin{equation}
    \frac{dL_r}{dm_r}=\epsilon_{nuc}+\epsilon_{grav}-\epsilon_{\nu}-\epsilon_{a}
    \label{energy_conservation_equation}
\end{equation}
where the energy production/consumption rates in the right-hand side are $\epsilon_{x}=\frac{dE_x}{dm_rdt}$ and $x$ account for the energy-balance contributions from nuclear reactions ($nuc$), changes of volume or internal energy ($grav$), neutrinos ($\nu$) and axions ($a$). Among the various contributions to the last term in this equation the one relative to the Primakoff process is proportional to the square of the axion-photon coupling, i.e., $g_{a\gamma}$, while others, such as those due to bremsstrahlung or Compton, are proportional to the square of the axion-electron coupling, i.e., $g_{ae}$ \citep[for review see][]{raffelt_book, caputoraffelt_2024,carenza_2025PhyR}.  
After the exhaustion of the central hydrogen, low-mass stars ($M<2$ M$_\odot$) move to the red giant branch (RGB), where the H burning is active in a thin shell located just outside a He-rich core. As the core mass increases temperature and density of the innermost zone increase as shown in Fig.\ref{fig:logT_profile} and Fig.\ref{fig:log_rho_profile}. In contrast, the H-rich envelope expands, so that  stars appear progressively brighter and cooler. Then, the RGB phase stops suddenly when core temperatures are high enough to trigger He burning. Since the central region of the core is efficiently cooled by the emission of neutrinos from plasmon decays, the conditions for He ignition are first attained  at $m_r\approx0.2 M_\odot$, where the density is \(\rho \simeq 10^6~\mathrm{g\,cm^{-3}}\) and the temperature is \(T \simeq 10^8~\mathrm{K} \approx 8.6~\mathrm{keV}\) (Fig.\ref{fig:star_rate_profile}). 
Within the core, the density is high enough for the electrons to become degenerate. Consequently, the degenerate equation of state triggers a thermal runaway known as the helium flash. Since the triple-$\alpha$ reaction rate is highly sensitive to both \(\rho\) and \(T\), an additional cooling channel, due to the thermal production of the axion, forces the core to grow more massive before reaching the conditions for He ignition. Thus the tip of the red giant branch (TRGB) would appear brighter (see Fig.\ref{fig:TRGB_RGBB_magnitude}). 

If axions are produced in the cores of RGB stars, the relevant density and temperature conditions imply that the plasma frequency becomes comparable to the axion energy. In this regime, the Primakoff process is strongly suppressed. Similarly, Compton scattering on degenerate electrons is disfavored due to Pauli blocking. Instead, axion production via bremsstrahlung may dominate, providing a non-negligible contribution to the stellar energy loss, see Fig.\ref{fig:star_rate_profile}. The TRGB luminosity can thus be used to constrain the strength of the axion--electron coupling, \(g_{ae}\), and currently provides some of the most stringent bounds \citep{straniero_2020,capozzi_2020}.  

Notably, particularly large axion-electron couplings ($g_{ae}\sim10^{-12}$) imply a very large core mass and hence very bright RGB stars. This triggers fast mass loss, causing the erosion of the H-rich envelope before reaching the He-flash. In that case, stars would move directly to the white dwarf locus, and no HB or AGB stars would be observed in GCs. This occurrence explains the saturation of the TRGB luminosity shown in Fig.~\ref{fig:TRGB_RGBB_magnitude}. Assuming, as usual, a Reimers mass-loss rate with $\eta=0.2$, we find that coupling values $g_{13}\gtrsim7.5$ can be ruled out. Although this limit depends on the assumed value of $\eta$, by considering a mass-loss rate in agreement with extant measurements for RGB stars \citep[][ and references therein]{mcdonald2015,tailo2020}, we find a very robust bound that is more than one order of magnitude smaller than the most stringent experimental constraints \citep{XENONnT}. This simple argument demonstrates the effectiveness of astrophysical bounds.

\section{Previous constraints on the axion-electron coupling from RGB stars}\label{sect:2}

\cite{viaux_2013} performed the first systematic use of the TRGB as proof of anomalous cooling in the core of RGB stars. They compared the I magnitude of the brightest RGB star of NGC 5904 (or M5) with stellar model predictions obtained under different assumptions about the axion-electron coupling. After a detailed analysis of the various uncertainties, they reported an upper bound of g$_{ae}<4.3\times 10^{-13}$ (95\%C.L.). Subsequently, \cite{capozzi_2020} following a previous claim by \cite{serenelli_2017}, applied an arbitrary  shift to the predicted $I$ magnitude of the RGB tip to account for a supposed inappropriate use of weak screening for the leading He-burning reaction, the triple$-\alpha$, in the PGPUC models used by \cite{viaux_2013}. 
Hence, based on the revised prediction and on the I magnitude of the NGC 5139 ($\omega$ Cen), they improved the bound to   g$_{ae}<1.3\times 10^{-13}$. However, as recently argued by \cite{valcarce_2025}, the PGPUC models actually adopt appropriate intermediate-screening prescriptions and, in turn, the upward shift of the theoretical TRGB luminosity is not justified. In addition, \cite{Soltis:2020gpl}, based on the parallax from Gaia EDR3, got a $\omega$ Cen distance similar to that adopted by \cite{capozzi_2020}, but with a doubled  error. For these reasons, the $g_{ae}$ bound reported \cite{capozzi_2020} should be relaxed upward.
Notably, the luminosity of a bright RGB star does not depend on its total mass, but, rather, on its He-core mass \citep{paczynsky_1970}. Therefore, the uncertainties affecting the physics of the H-rich envelope, e.g., those related to the mixing-length parameter, the mass-loss rate, the atomic and molecular opacity or the adopted surface boundary conditions, produce negligible effects on the predicted RGB tip luminosity. Conversely, these uncertainties impact the predicted effective temperature ($T_{eff}$) and, in turn, the resulting $I$ or $V$ mag predictions, given the dependence of the bolometric correction on $T_{eff}$. Indeed, luminosity is a far more robust stellar-model prediction than the effective temperature. To mitigate this problem, a different approach was followed by \cite{straniero_2020} \citep[also see][]{Straniero_2018}, where 22 GCs have been considered by combining optical and near-infrared data from public archives of ground and space based telescopes (HST). Thanks to this multi-wavelength global analyses, they derived the bolometric magnitude of the brightest RGB stars of each cluster, a quantity that can be directly compared to the corresponding luminosity predicted by stellar models accounting for the additional axion cooling.  In this manner, the TRGB luminosity of each GC is determined entirely by observables, specifically magnitudes, colors, distance, and interstellar extinction, thereby bypassing the need for theoretical effective temperatures predicted by stellar models.
First, they applied this  method to three globular clusters (hereafter GCs), namely,  NGC 362, NGC 5904 (M5) and NGC 104 (47 tuc) finding  $g_{ae}/10^{-13}$ bounds of 1.37, 2.30 and 1.87, respectively (95 \% C.L.). Moreover, the global likelihood analysis, as obtained combining the RGB-tip measurements of 22 GC, fixed the $g_{ae}$ bound to $<1.48\times 10^{-13}$, further reduced  down to $0.95\times 10^{-13}$ \citep[see the review paper by][] {carenza_2025PhyR}, after adopting updated distances \citep{baumgardt_2021MNRAS}.

We recall that complementary constraints on the axion--electron coupling can be obtained from white-dwarf (WD) cooling, where axion bremsstrahlung provides an additional energy-loss channel~\cite{Isern:2008nt, Isern:2018uce, MillerBertolami:2014rka}. Using the 100-pc Gaia DR3 WD luminosity function and updated population-synthesis and cooling models, \citet{Alberino:2026yxi} obtained the bound ($g_{ae}<1.68\times10^{-13}$) at $95\%$ C.L., finding no evidence for the mild preference for additional cooling reported in some earlier analyses. A stronger limit, ($g_{ae}\leq0.81\times10^{-13}$), has recently been reported by \citet{Fleury:2025ahw} from the WD cooling sequence of 47 Tuc. The latter result, however, appears more sensitive to assumptions entering the WD and cluster-population modeling. In particular, the constraint relies on informative priors on the WD mass and hydrogen-envelope thickness derived within a specific cooling framework, while other common theoretical uncertainties are not simultaneously propagated. Moreover, the WD birthrates preferred by the fit differ from estimates based on the rate at which RGB stars leave the main sequence \citep{Goldsbury:2016ndn}; notably, the preferred rate is lower in the inner WFC3 field and higher in the outer ACS field. \citet{Fleury:2025ahw} themselves suggest that this discrepancy may reflect the dynamical redistribution of WDs within the cluster, an effect not included in their likelihood. We regard this constraint as complementary but comparatively model dependent.

\section{A new method to constrain the axion-electron coupling}\label{sect:3}

\begin{figure}
    \centering
    \includegraphics[width=\linewidth]{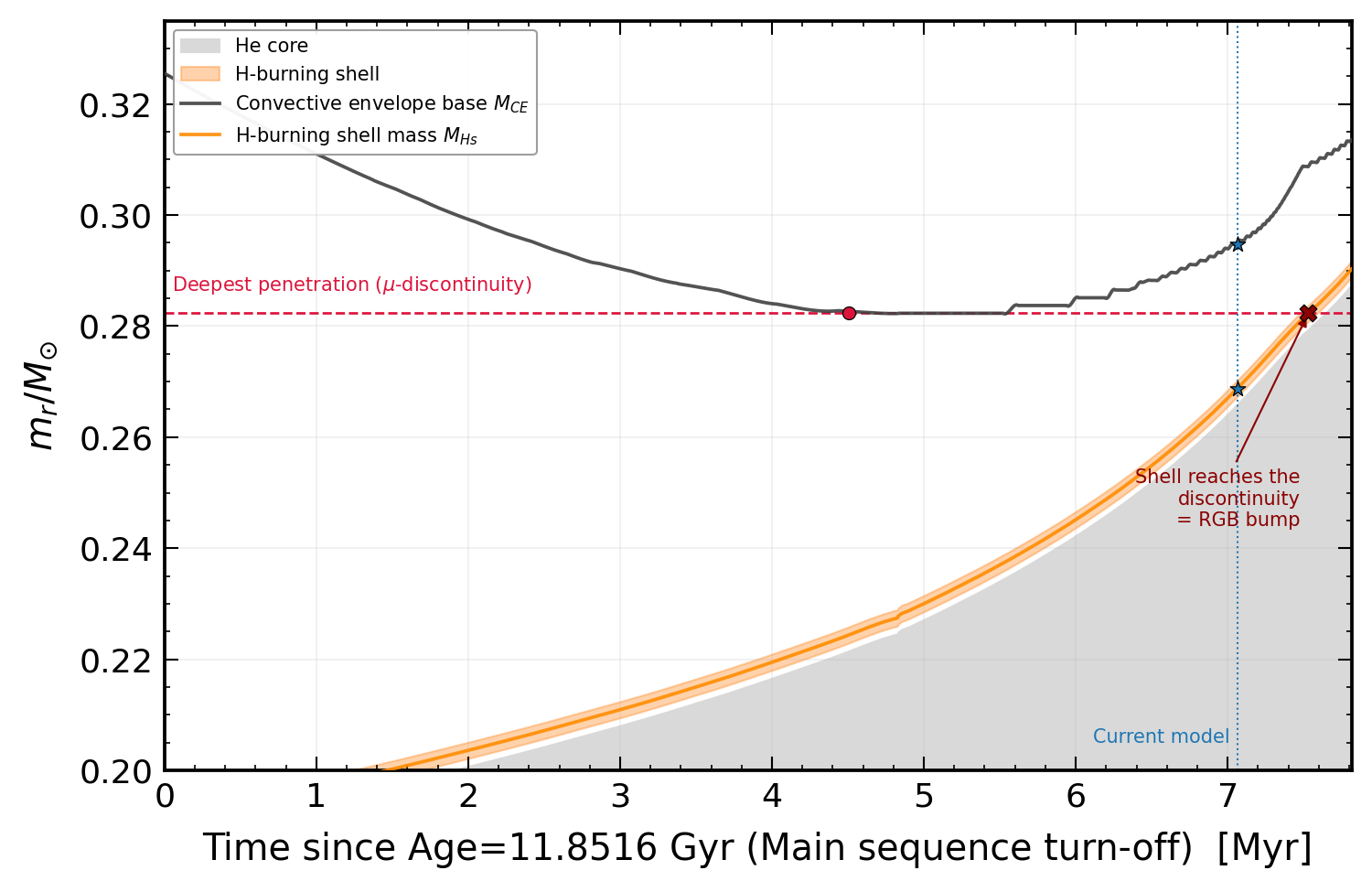}
    \caption{Time evolution of the position of the base of the convective shell, the He core and the H-burning shell. The RGBB happens when the burning shell reaches the point of deepest penetration of the envelope.}
    \label{fig:RGBB_mechanism_timeline}
\end{figure}

\begin{figure}
    \centering
    \includegraphics[width=\linewidth]{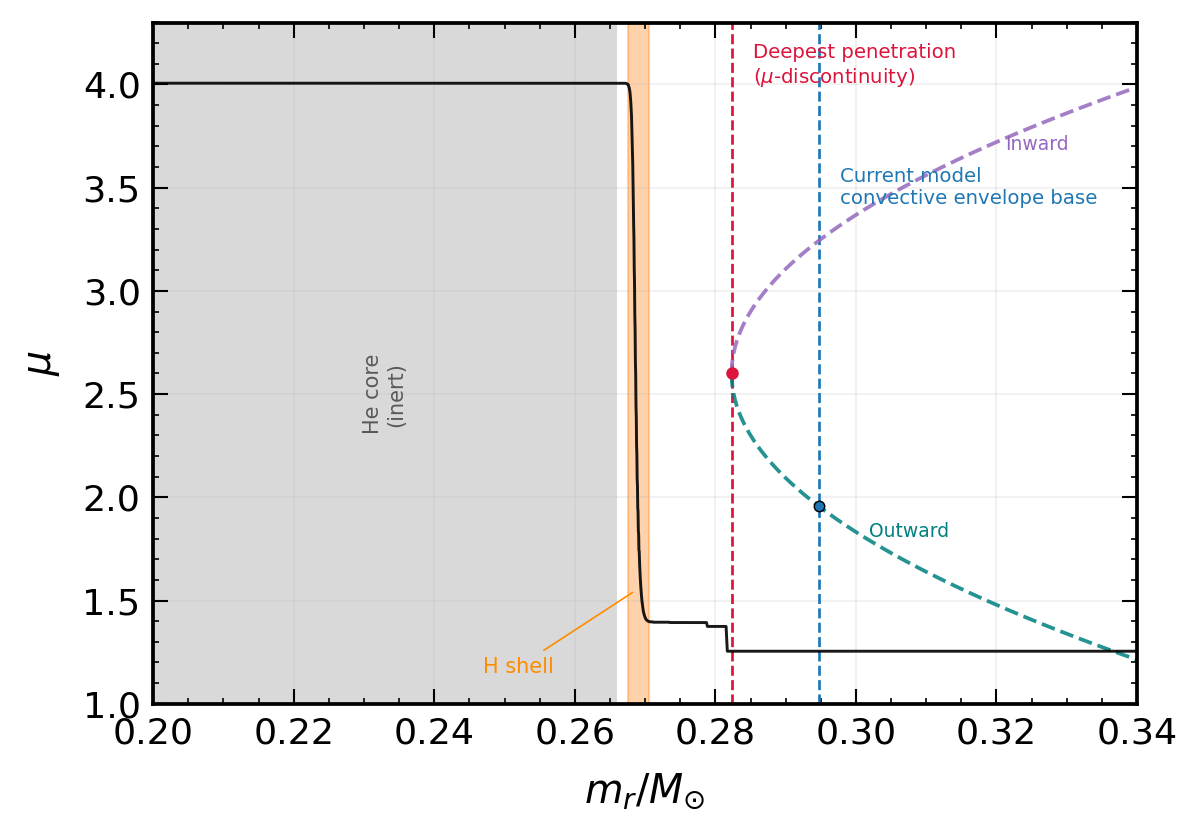}
    \caption{Mean atomic weight $\mu$ profile for a RGB star before the RGBB. The dashed lines show the evolution of the base of the convective envelope: the envelope moves inward (purple line), reaches a deepest point before (red line) leaving a discontinuity in the atomic weight profile, before retreating (green line). The blue line shows the position of the envelope base of the current model. The He core and H-burning shell are also shown.}
    \label{fig:RGBB_mechanism}
\end{figure}

As the difference between standard theoretical predictions and the observed RGB tip luminosity is minimal, the derived bound is primarily controlled by the overall error budget. For this reason, statistical and systematic errors affecting the measured stellar luminosity should be carefully estimated, as well as all the possible uncertainties affecting the corresponding theoretical predictions.
Among the uncertainties affecting the derived bound, a major contribution to the error budget is still given by the distance determinations. Also the bolometric corrections  used to transform the measured magnitude into the absolute stellar luminosity could hide systematic uncertainties not properly taken into account. 
The alternative method proposed here aims to reduce, or even eliminate, many of these systematic errors. 
The core idea is to constrain $g_{ae}$  through the TRGB brightness relative to an HR diagram reference point that is only marginally affected by axion induced cooling, specifically, the RGB bump (RGBB). First predicted by \cite{Thomas_1967} and \cite{iben_1968}, it happens when the H burning shell reaches the point of deepest penetration of the convective envelope (Fig.\ref{fig:RGBB_mechanism_timeline}). There, the envelope has dredged up unprocessed H from the surface, leaving a discontinuity in the mean atomic weight (Fig.\ref{fig:RGBB_mechanism}). When crossing this discontinuity, the increase in luminosity of the ascending RGB momentarily stalls and decreases (Fig.\ref{fig:diff_conf}). Because the star has to cross this region of the CMD three time, thus spending more time there, RGB stars accumulate in this region and appear in the luminosity function of the RGB as a luminosity "bump".

\begin{figure}
    \centering
    \includegraphics[width=\linewidth]{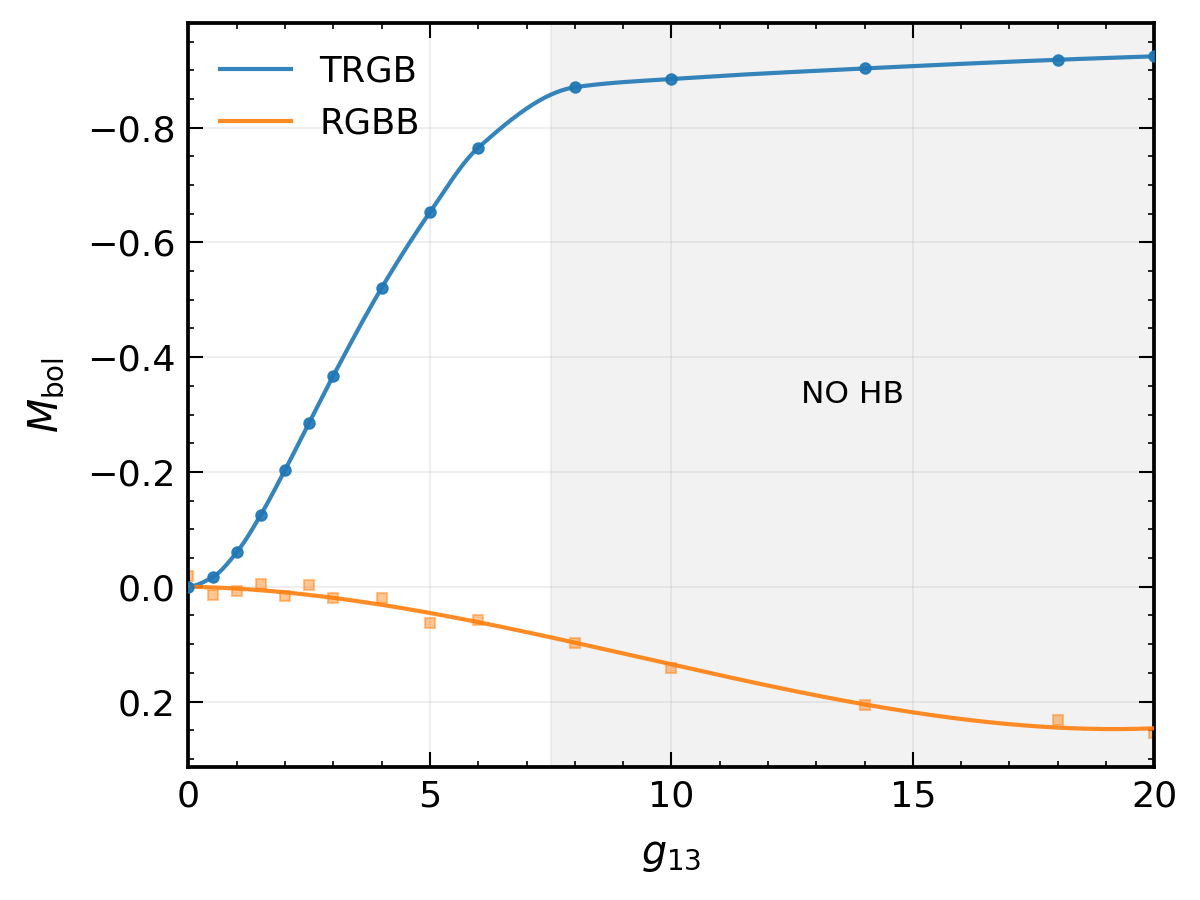}
    \caption{$M_{\rm TRGB}$ and $M_{\rm RGBB}$ relative to their standard-physics values ($g_{13}=0$), for models with $Y=0.25$, $Z=2.0\times10^{-3}$ and mass $M=0.86\,M_{\odot}$. Over the range of $g_{13}$ explored in this work, the RGBB magnitude is essentially insensitive to the axion-electron coupling. For large values of the axion-electron coupling ($g_{13} > 7.5$), the high luminosity reached by the star drives a substantial increase in mass loss, resulting in the loss of almost the entire envelope before reaching the conditions for helium ignition in the core. Consequently, upon reaching this luminosity, stars exit the RGB prior to the He-flash, bypassing later evolutionary stages, specifically the HB and AGB. The shaded area marks the forbidden region due to the disappearance of these stellar evolution phases.
    }
    \label{fig:TRGB_RGBB_magnitude}
\end{figure}

In practice, we propose to use the bolometric-magnitude difference between the TRGB and the RGBB, $\Delta M_{TRGB-RGBB}$. 
As illustrated in Fig.\ref{fig:TRGB_RGBB_magnitude}, the RGBB is practically insensitive to the axion-electron coupling for $g_{13}<5$, so that the expected change in  $\Delta M_{TRGB-RGBB}$ is identical to the shift in the tip bolometric magnitude.
This approach makes the bound independent of the adopted distance scale, extinction coefficient and of the zero-point of the bolometric corrections.
However, the RGBB magnitude is more sensitive to variations of the cluster parameters, specifically, the age and He content, and their variations within reasonable ranges of values should be carefully taken into account (see Tab. \ref{tab:clusterparameters}).

In the past, the capability of extant stellar models in reproducing the RGBB luminosity of GCs has been questioned \citep{fusipecci_1990,Cassisi-Innocenti-Salaris,2010ApJ...712..527D}. According to these studies, it appears that the theoretical predictions for the V magnitude of the RGBBs are systematically brighter than the observed one, at least, for metallicities lower than [$\rm{[Fe/H]}<-1.7$, while for higher metallicity, the discrepancy is confined within the current uncertainties. If this is the case, the constraint to the axion-electron coupling we are proposing would be affected by a systematic error causing an overestimation of the $g_{ae}$ bound. For this reason, in this first work  we will limit our analysis to GCs with a sufficiently high metallicity, namely NGC 5904 and NGC 362 ($\rm{[Fe/H]}\approx -1.3$), and NGC 104 ($\rm{[Fe/H]}\approx -0.7$). Anyway, the occurrence of a possible systematic error in the theoretical estimation of the RGBB luminosity for these three clusters will be carefully checked and discussed in Sec. \ref{sect:RGBB_reliability}.

\section{RGB Bump and Tip in NGC 104, NGC 362 and NGC 5904}\label{sect:observations}
  
In this section, we describe the methodology used to determine the bolometric magnitudes of the RGB bump and tip for three GCs, namely: NGC 104 (47 Tuc), NGC 362, NGC 5904 (M5). The input data are from specific catalogs of  Johnson-Cousins V- and I-bands magnitudes whose construction is fully described in section 3.1 of \citet{straniero_2020}. 
In order to get a large and complete sample of RGB stars, for each cluster, the adopted strategy was to combine different photometric catalogs characterized
by complementary properties. In particular, high angular resolution photometric data from the ACS globular cluster survey \citep{Sarajedini} were used to resolve the crowded central regions, while 
to sample the entire radial extension of the clusters, photometric data from large field-of-view (FOV) ground-based facilities have been used \citep{Stetson}.
In the case of NGC 362, since the ground-based catalog was not complete as for
NGC 5904 and NGC 104, additional observations from the FORS2 data (ESO archive program 60.A-9203) were also used.
In addition, particular attention has been paid to decontaminate the RGB photometric samples from background field stars and from the presence of AGB stars \citep[see][for more details]{straniero_2020}.

Tab. \ref{tab:clusterparameters} summarizes the main cluster parameters. 
[Fe/H] is from \cite{carretta_2009AA}, while [M/H] is obtained by means of equation 2 in \cite{straniero_2020}, assuming a mean [$\alpha$/Fe]$=0.4\pm0.1$. We take a uniform distribution of GC age between 11 Gyr and 13 Gyr as a  rather conservative prior, and we assume a helium abundance sampled over a uniform distribution $0.245-0.255$ around the primordial abundance one \citep{Fields_2020}.
The true distance moduli are derived from the distances obtained by \cite{baumgardt_2021MNRAS}. They are based on a combination of Gaia EDR3, HST and literature data. E(B-V) values  are from \cite{Harris_2010}.  
Among these cluster parameters, only age and metallicity play a role in our determination of the axion-electron coupling constraint.  Indeed, they will be used to select the appropriate evolutionary tracks for the comparisons with the RGB photometric data. Instead, distance and reddening, which are required to convert apparent magnitudes into absolute magnitudes, cancel out when calculating the luminosity differences between TRGB and RGBB, and, therefore, do not affect our results. Actually, the reddening will be used to convert the apparent V-I color into (V-I)$_0$, a quantity used as input for calculating the bolometric corrections. However, the error associated to the reddening is generally small compared to the overall uncertainty of the bolometric corrections.

\begin{table*}[t]
\centering
\caption{Adopted cluster parameters.}
\label{tab:clusterparameters}

\begin{tabular}{lccc}
\toprule
   & NGC 104 & NGC 362 & NGC 5904 \\
\midrule
\multicolumn{4}{l}{parameters used to derive the axion-coupling bound:} \\
$\rm{[Fe/H]}$ & $-0.76\pm0.02$ & $-1.30\pm0.04$ & $-1.33\pm0.02$ \\
${\rm[M/H]}$ & $-0.46\pm0.09$ & $-1.00\pm0.09$ & $-1.03\pm0.09$ \\
Y & 0.245-0.27 & 0.245-0.27 & 0.245-0.27 \\
age (Gyr) & 11-13 & 11-13 & 11-13  \\
\midrule
\multicolumn{4}{l}{parameters used only for the calculation of the bolometric corrections:} \\
E(B-V)  & $0.04\pm0.1$   & $0.05\pm0.1$   & $0.03\pm0.1$   \\
\midrule
\multicolumn{4}{l}{parameters used only for the comparison among theoretical and observed RGBB luminosities:} \\
(m-M)$_0$ & $13.281\pm0.015$ & $14.753\pm0.023$ & $14.387\pm0.017$ \\
\bottomrule
\end{tabular}

\tablefoot{ The total metallicity, [M/H], is obtained from the [Fe/H] reported by \cite{carretta_2009AA} and assuming $[\alpha/\text{Fe}]=0.4\pm0.1$ (see text). No priors have been assumed for Y amd age in the bound analysis. Reddenings are from \cite{Harris_2010}, while distance moduli from \cite{baumgardt_2021MNRAS}. }
\end{table*}

\subsection{Finding the RGB bump}\label{sec:obsv_RGBB}

\begin{figure*}[t]
    \centering

    \begin{minipage}{0.48\textwidth}
        \centering
        \includegraphics[width=\linewidth]{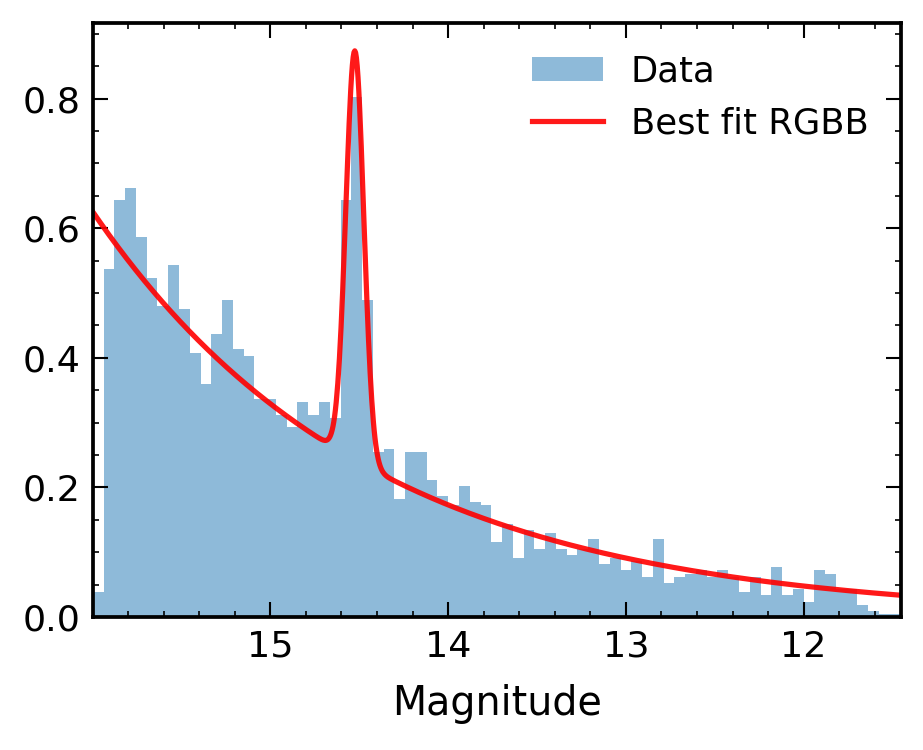}
    \end{minipage}
    \hfill
    \begin{minipage}{0.48\textwidth}
        \centering
        \includegraphics[width=\linewidth]{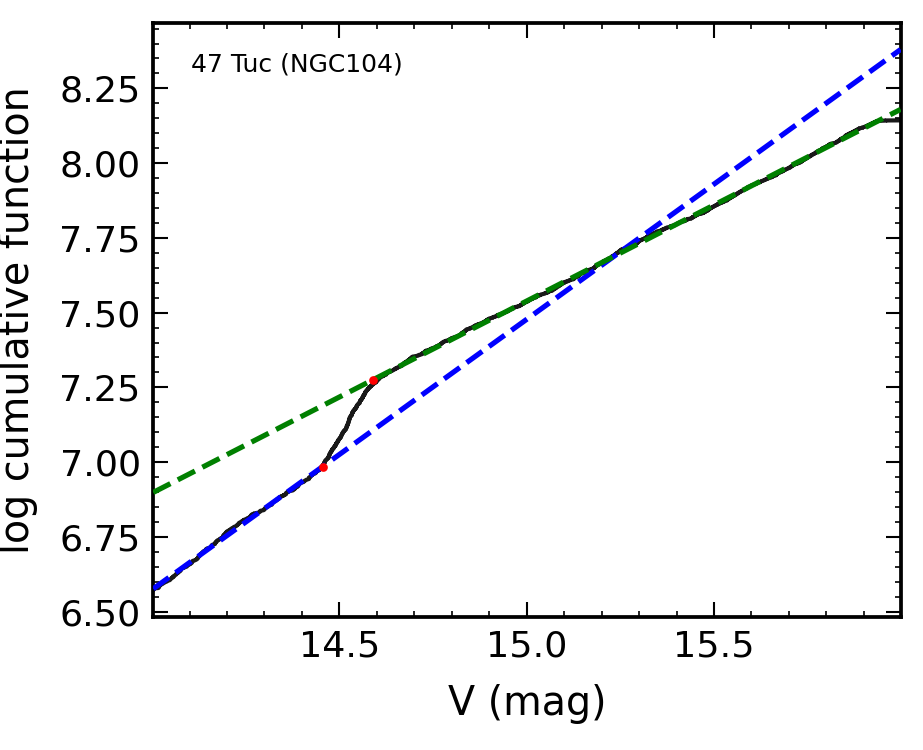}
    \end{minipage}

    \vspace{2mm}

    \begin{minipage}{0.48\textwidth}
        \centering
        \includegraphics[width=\linewidth]{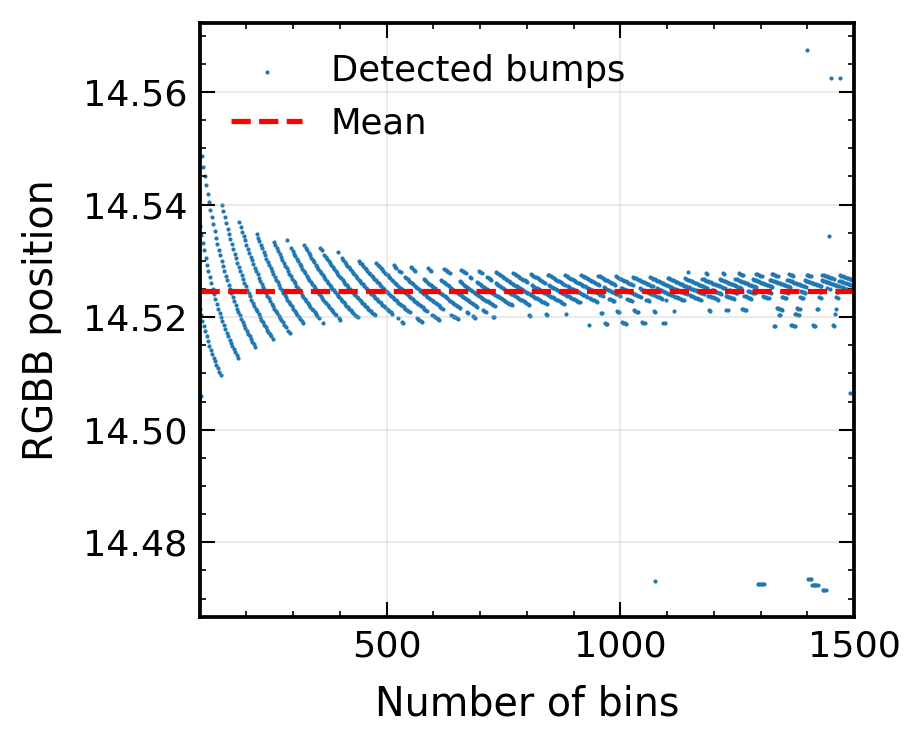}
    \end{minipage}
    \hfill
    \begin{minipage}{0.48\textwidth}
        \centering
        \includegraphics[width=\linewidth]{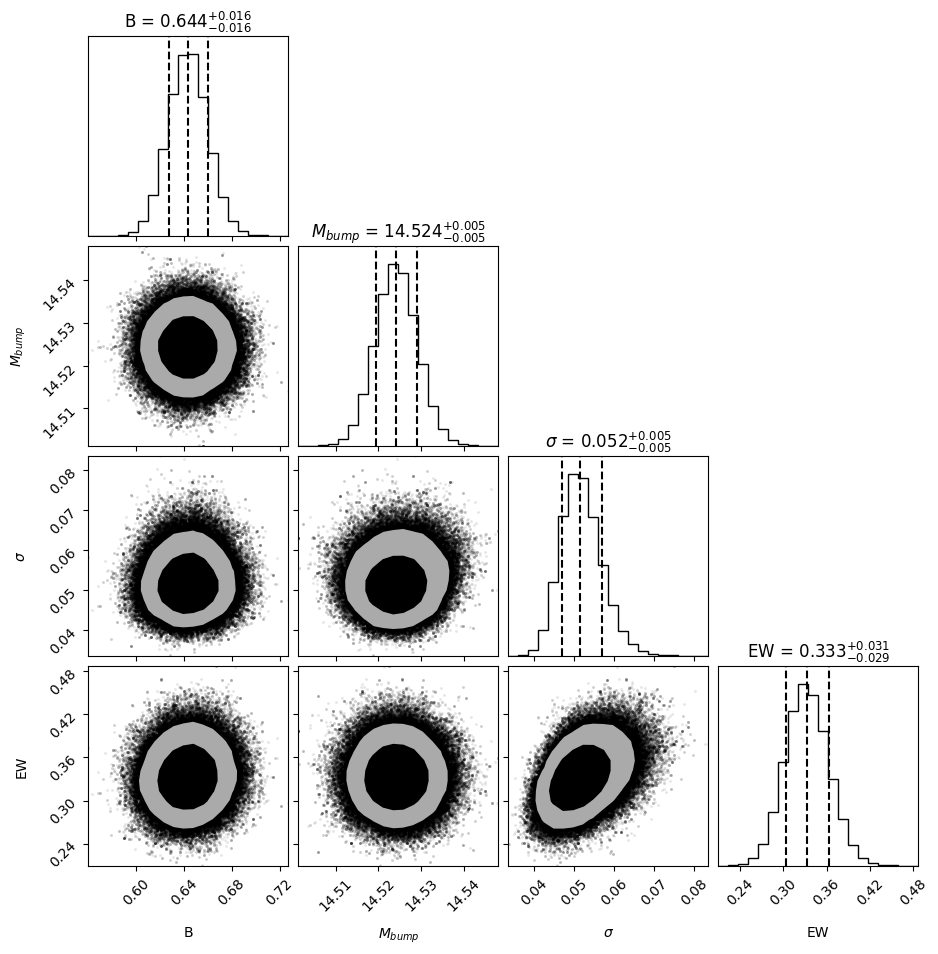}
    \end{minipage}

    \caption{RGBB selection method comparison for the V magnitude of 47tuc: Fusi-Pecci method (top-right), bump stability (bottom-left) and corner plot of the MCMC (bottom-right). The histogram as well as the best fit from the MCMC is shown on the top-left corner. All three methods gives similar results.}
    \label{fig:RGBB_selection}
\end{figure*}

In order to determine the luminosity of the RGB bump and its uncertainties 
we apply three distinct methods. The first follows the procedure described by \cite{fusipecci_1990}. This method uses the logarithm of the cumulative luminosity function (CLF), where the region around the RGBB appears as two linear segments with a slope break between them. We fit these three portions and take the central value of the break as the RGBB, with the associated uncertainty given by the width of the slope change (see Fig. \ref{fig:RGBB_selection}) .

The second method relies on the stability of the RGBB position in the luminosity function (LF), when the bin sizes are varied across a reasonable range. The location of the LF relative maximum typically converges to one or two closely spaced values. We then calculate the mean of these stable positions, excluding outliers, specifically points lying beyond $1.5 \times \mathrm{IQR}$ from the first and third quartiles, and estimate the uncertainty using the standard deviation.

The third method is the one originally introduced by \cite{nataf_2013ApJ}. It is based on a specific assumption about the shape of the RGB LF, namely:
  
\begin{equation}
\begin{split}
N(M) = A \times &\Bigg\{ \exp\left[B(M-M_{RGBB})\right] \\
&+ \frac{EW_{RGBB}}{\sqrt{2\pi}\sigma_{RGBB}}
\exp\left[-\frac{(M-M_{RGBB})^2}{2\sigma_{RGBB}^2}\right]
\Bigg\}
\end{split}
\end{equation}
where M is a V or I magnitude, $M_{RGBB}$ the RGBB magnitude in corresponding color, $\sigma_{RGBB}$ its width, $EW_{RGBB}$ a normalization factor that gives the number of stars\footnote{Actually the number of stars within the bump is given by $A\times EW_{RGBB}$} within the bump. B drives the overall exponential decrease of the LF while A is an overall normalization factor related to the total number of observed RGB stars. B,$M_{RGBB}$,$\sigma_{RGBB}$ and $EW_{RGBB}$ are determined by fitting the observed LF using a Markov Chain Monte Carlo (MCMC) method, while A is computed to match the number of stars. 
For a full discussion on the method robustness, see \citet{nataf_2013ApJ}.

The resulting magnitudes and colors are compared in Tab. \ref{tab:bumpmag} and \ref{tab:bumpcolor} and illustrated in Fig.\ref{fig:RGBB_selection} for the V magnitude of 47TUC. Available values previously obtained by \cite{fusipecci_1990} and \cite{nataf_2013ApJ} are also reported. Within the uncertainties, the three methods provide very similar results, all in excellent agreement with those obtained in previous studies. In the following, we will adopt the weighted mean of the  V and I values obtained with the three methods. 

\begin{table*}[t]
\centering
\caption{Apparent RGBB V and I mag of NGC 104, NGC 362 and NGC 5904, as obtained by means of the three methods described in section \ref{sec:obsv_RGBB}.}
\label{tab:bumpmag}
\resizebox{\textwidth}{!}{%
\begin{tabular}{lcccccc}
\toprule
 & \multicolumn{2}{c}{NGC 104} & \multicolumn{2}{c}{NGC 362} & \multicolumn{2}{c}{NGC 5904} \\
\midrule
     & V & I & V & I & V & I \\
\midrule
Slope break & $14.522\pm0.067$ & $13.487\pm0.063$  & $15.398\pm0.071$ & $14.415\pm0.058$ & $14.968\pm0.054$ & $13.968\pm0.055$ \\
Bump stability &  $14.525\pm0.003$ & $13.488\pm0.010$ & $15.394\pm0.008$ & $14.393\pm0.009$ & $14.970\pm0.017$ & $13.972\pm0.032$ \\
MCMC & $14.524\pm0.005$ & $13.489\pm0.005$ & $15.402\pm0.008$ & $14.418\pm0.007$ & $14.969\pm0.008$ & $13.972\pm0.006$ \\
\midrule
weighted mean & $14.525\pm0.003$ & $13.489\pm0.004$ & $15.398\pm0.006$ & $14.409\pm0.006$  & $14.969\pm0.006$ &  $13.972\pm0.003$ \\
\midrule
N2013 & $14.507\pm0.005$ & & $15.399\pm0.007$ & & $14.963\pm0.009$ & \\
FP1990 & $14.55\pm0.05$ &  & & & $15.05\pm0.05$ & \\
\bottomrule
\multicolumn{7}{l}{{\it Note}: N2013 - \cite{nataf_2013ApJ}, FP90 - \cite{fusipecci_1990}}
\end{tabular}%
}

\end{table*}

\begin{table*}[t]
\centering
\caption{Apparent RGBB V-I colors of NGC 104, NGC 362 and NGC 5904, as obtained by means of the three methods described in section \ref{sec:obsv_RGBB}.}
\label{tab:bumpcolor}
\begin{tabular}{lccc}
\toprule
   & NGC 104 & NGC 362 & NGC 5904 \\
\midrule
Slope break & $1.04\pm0.09$ & $0.98\pm0.09$ & $1.00\pm 0.08$ \\
Bump stability & $1.04\pm0.01$ & $1.00\pm0.01$ & $1.00\pm0.04$ \\
MCMC & $1.035\pm0.007$ & $0.984\pm0.010$ & $0.998\pm0.010$ \\
weighted mean & $1.037\pm0.006$ & $0.992\pm0.007$ & $0.998\pm0.010$ \\ 
\midrule
\cite{nataf_2013ApJ} & $1.03$ & $0.98$ & $1.01$ \\
\bottomrule
\end{tabular}
\end{table*}

\subsection{Finding the RGB tip}
\begin{figure}[htbp]
    \centering
    \includegraphics[width=0.8\columnwidth]{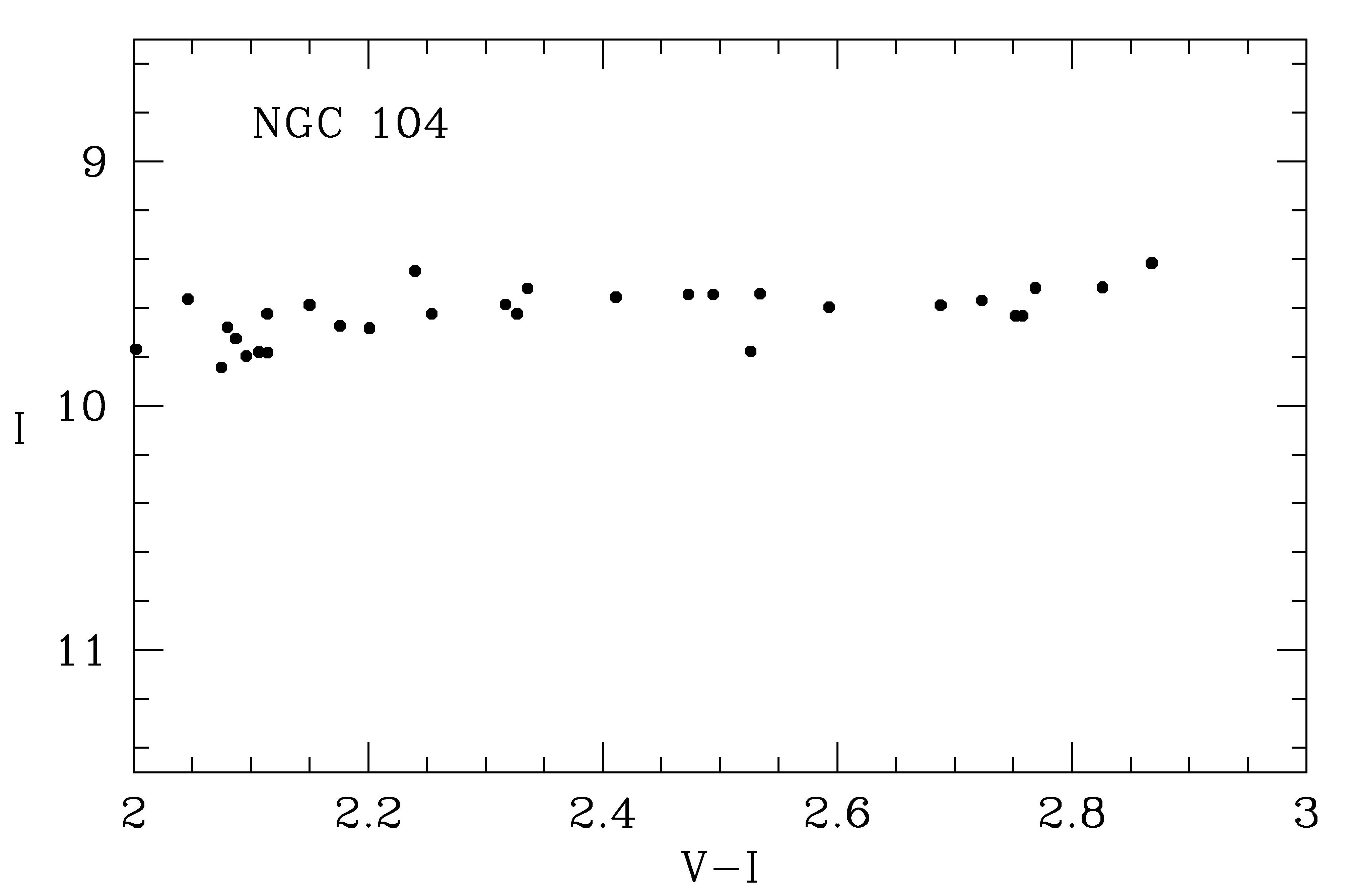}
    \vspace{2mm} 
    \includegraphics[width=0.8\columnwidth]{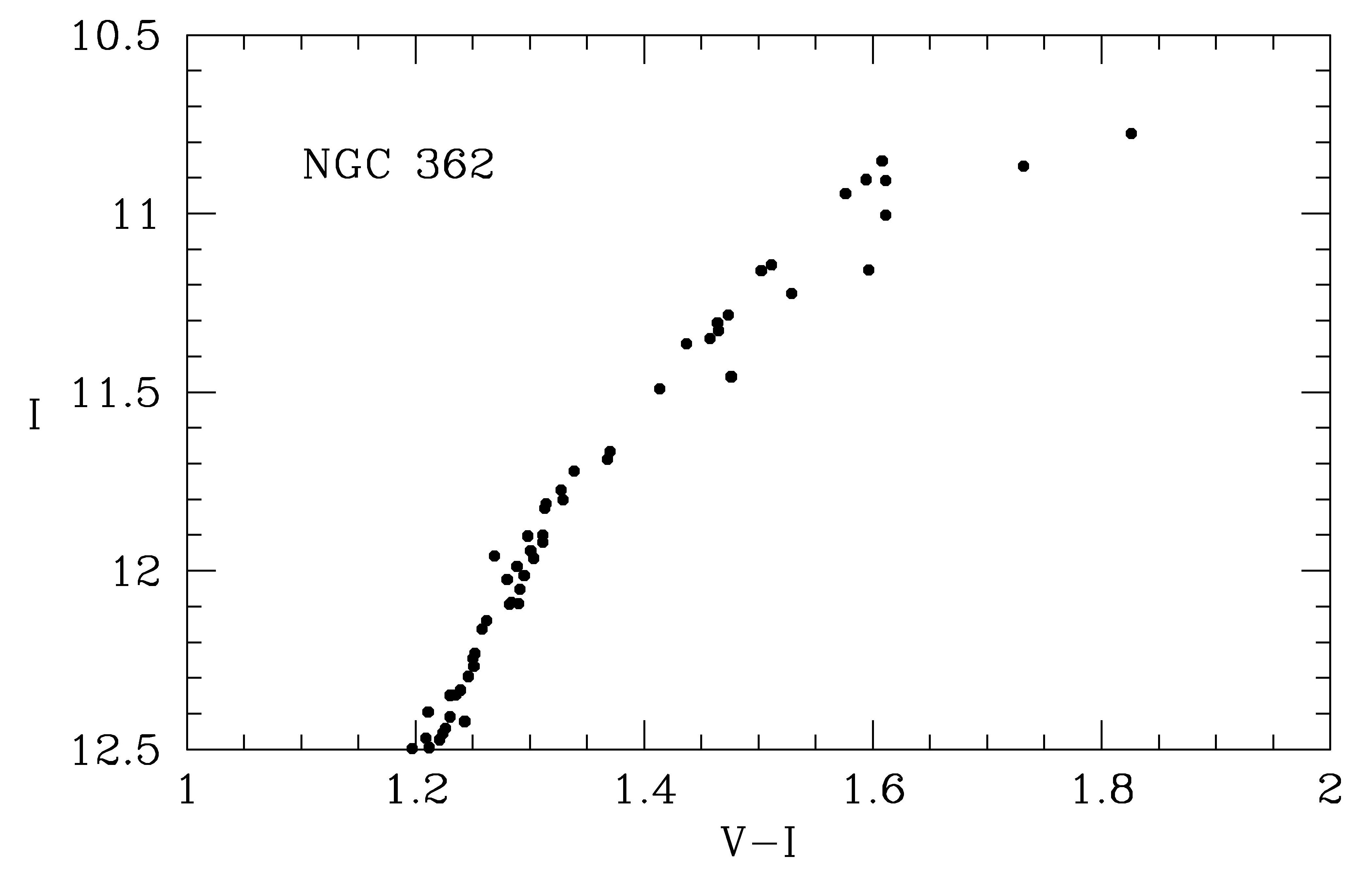}
    \vspace{2mm} 
    \includegraphics[width=0.8\columnwidth]{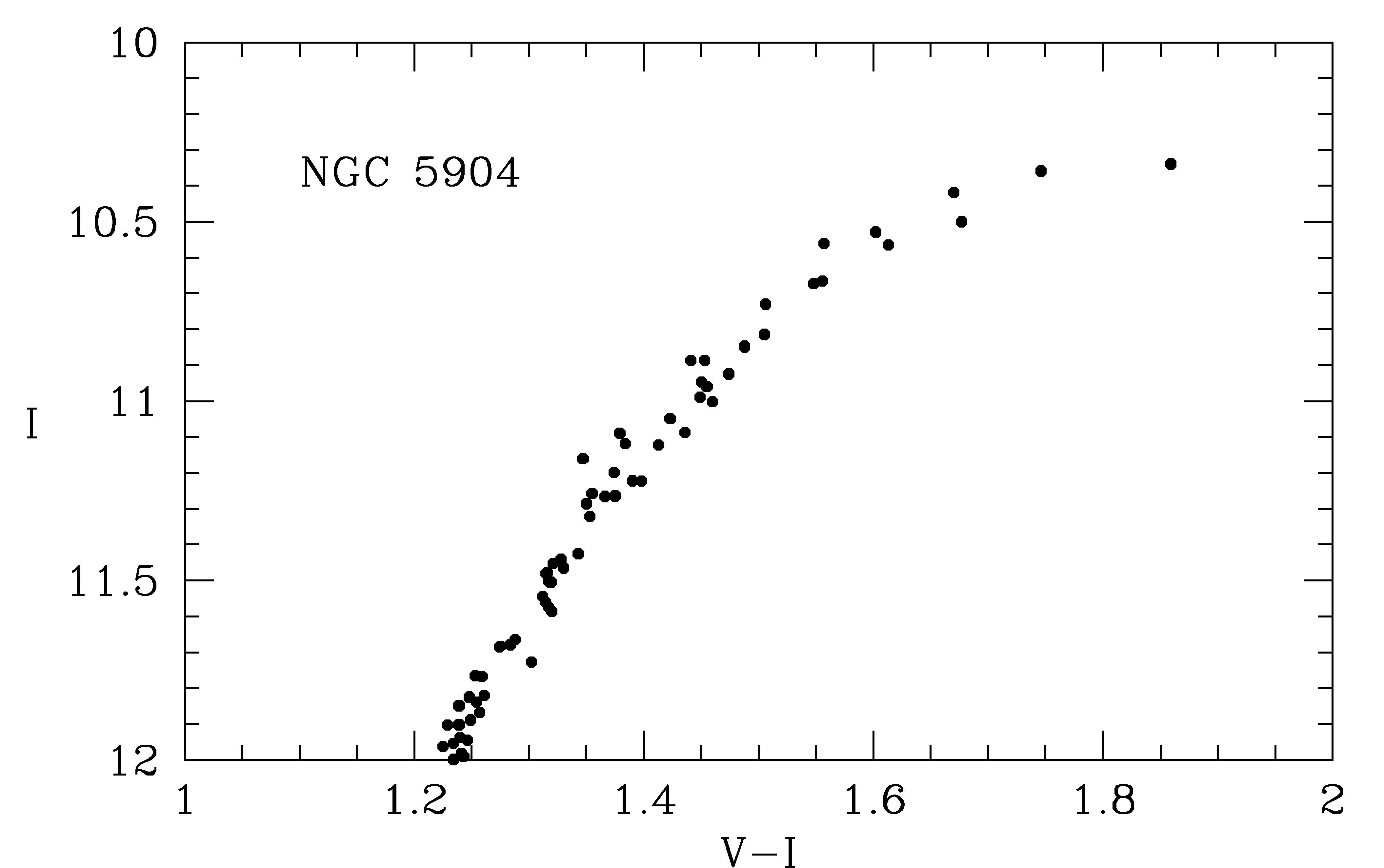}
    
    \caption{Color-magnitude diagrams of the brightest RGB portions for NGC 104, NGC 362 and NGC 5904.}
    \label{fig:cmds}
\end{figure}

To determine the RGB tip, we follow the procedure described in \cite{straniero_2020}. The brightest portion of the RGB color-magnitude diagrams are shown in Fig. \ref{fig:cmds}. Firstly, we eliminate residual contamination from field or AGB stars, by searching for stars that clearly deviate from the fiducial RGB sequence. Next, we identify the stars with the reddest $V-I$ color indices. These are indeed the brightest stars, as a larger $V-I$ color implies a greater bolometric correction.  For each cluster, Tab. \ref{tab:tipmag} reports the coordinates (J2000) of the brightest RGB star and its V and I apparent magnitudes. 
The main difficulty is to determine the TRGB magnitudes with a limited sample of stars in the brightest part on the RGB. Therefore, to evaluate the separation between the magnitude of the brightest star ($m_{BS}$) and that of the TRGB ($m_{tip}$), we generate $5\times10^4$ synthetic color-magnitude diagrams (SCMDs) for each cluster. Each SCMD contains the same number of stars in the brightest 2.5 bolometric magnitudes of the RGB as in the observed photometric catalogs. Defining the correction $\delta m = m_{BS} - m_{tip}$, we obtain  the statistical distribution of the correction to be applied to the brightest star (see Fig\ref{fig:TRGB correction} for 47TUC). The SCMDs are generated using the cluster parameters from Tab. \ref{tab:clusterparameters}. We confirmed that our results are robust against reasonable variations in age and metallicity, with similar findings for both V and I bands.
The results for the three clusters are summarized in Tab. \ref{tab:tipmag}. In the subsequent Monte Carlo analysis, we will extract the correction from the corresponding distributions.

\begin{table*}
\centering
\caption{Quantities used for the determination of the TRGB apparent brightness and color, namely: right ascension (RA) and declination (Dec) of the RGB stars with the largest V-I; apparent V, I and V-I of the brightest RGB stars; number of RGB stars within 2.5 mag from the tip.}
\label{tab:tipmag}
\resizebox{\textwidth}{!}{%
\begin{tabular}{lccc}
\toprule
 & NGC 104 & NGC 362 & NGC 5904 \\
\midrule

Brightest star coordinates (RA/DEC) & (6.0020833/-72.1240833) & (15.8410907/-70.9055993) & (229.6501803/2.1104192) \\
\midrule
V (brightest star) & $12.285 \pm 0.007$ & $12.602 \pm 0.008$ & $12.198 \pm 0.010$ \\
I (brightest star) & $9.417 \pm 0.015$ & $10.776 \pm 0.008$ & $10.339 \pm 0.008$ \\
V-I (brightest star) & $2.868 \pm 0.016$ & $1.826 \pm 0.011$ & $1.859 \pm 0.013$ \\
\midrule
Number of bright RGB stars & 108 & 88 & 122 \\

\bottomrule
\end{tabular}%
}
\end{table*}

\subsection{TRGB and RGBB bolometric magnitudes}\label{sec:bolometric}

In order to compare with model predictions, we transform the TRGB and the RGBB apparent V or I magnitudes into bolometric magnitudes.   We calculate the bolometric corrections ($BCs$), using as inputs [M/H], $(V-I)_0$ and and the gravity ($\log g$). Therfore we need to correct the apparent $V-I$ color  for extinction. As usual, extinction coefficients are given by $A_V=3.1\times E(B-V)$ and $A_I=0.6\times A_V$. We use  reddenings and the metallicities listed in Tab. \ref{tab:clusterparameters}. Note that for RGB stars, the $BCs$ are marginally affected by the adopted gravity. Basing on typical luminosities and radii of RGB stars with metallicity [M/H] in the range -1 and -0.5, we adopt $\log g=2.0$ and $0.0$, for the RGBB and the TRGB, respectively.
Three sets of BCs are considered: specifically, those based on ATLAS9 atmospheric models \citep{castelli_2003}, those based on MARCS atmospheric models \citep{MARCS}, and the empirical ones by \cite{worthey_2011} (hereafter WL2011). Note that the ATLAS9 models do not cover the particularly cool atmospheres of stars near the RGB tip of NGC 104. The MARCS models, on the other hand, were specifically designed for the study of cool stars such as RGBs \citep{MARCS}. The different BCs introduce an important source of uncertainty that has been overlooked in the past. For $\text{[M/H]} = -1$ and $V-I = 1.85$, typical of the brightest RGB stars, we found $BC_V = -1.56$, $-1.45$, and $-1.39$ for ATLAS9, MARCS, and WL2011, respectively, while for the RGBB ($V-I = 1.0$), we found $-0.45$, $-0.36$, and $-0.36$, respectively. These differences among the three bolometric correction sets are not negligible for the purposes of determining the axion bound.
The discrepancy have multiple origins. Synthetic models (ATLAS9, MARCS) differ significantly in their treatment of atmospheric geometry. Specifically, ATLAS9 uses a 1D plane-parallel approximation, whereas MARCS incorporates 1D spherical geometry, which is essential for extended giant envelopes. Furthermore, variations in molecular opacities and line blanketing (from species such as TiO, CO, and H$2$O) alter the predicted flux redistribution from optical to infrared wavelengths. In contrast, empirical relations like WL2011 rely on observational calibrations, which carry uncertainties in reddening, $T{\text{eff}}$ scales, and sample coverage. These discrepancies mainly impact the derived bolometric luminosity of the RGB tip, whch is highly sensitive to small variations of the $V-I$ color.  
As pointed out by \citet{straniero_2020}, the derivation of the bolometric magnitude of stars close to the TRGB is significantly more reliable when using the near-infrared color $J-K$ and its corresponding correction $BC_K$. This advantage arises because $BC_K$ exhibits a remarkably flat behavior and weak dependence on $J-K$, making it exceptionally resilient to color uncertainties. In contrast, small changes in $V-I$ significantly affect the resulting $BC_V$ and, similarly, $BC_I$. Moreover, near-IR magnitudes are less affected by uncertainties in molecular line blanketing and interstellar extinction. Consequently, relying on $BC_K$ drastically mitigates systematic uncertainties in the determination of the TRGB $M_{\text{bol}}$ and the resulting axion bounds. Nevertheless, since the primary goal of the present work is to introduce the new method for constraining the axion coupling, we restrict our application to the available $V$ and $I$ photometric catalogs detailed at the start of Section \ref{sect:observations}, fully aware of the intrinsic limitations discussed here.

\section{Stellar models}\label{sect:models}

\begin{figure}
    \centering
    \includegraphics[width=0.8\columnwidth]{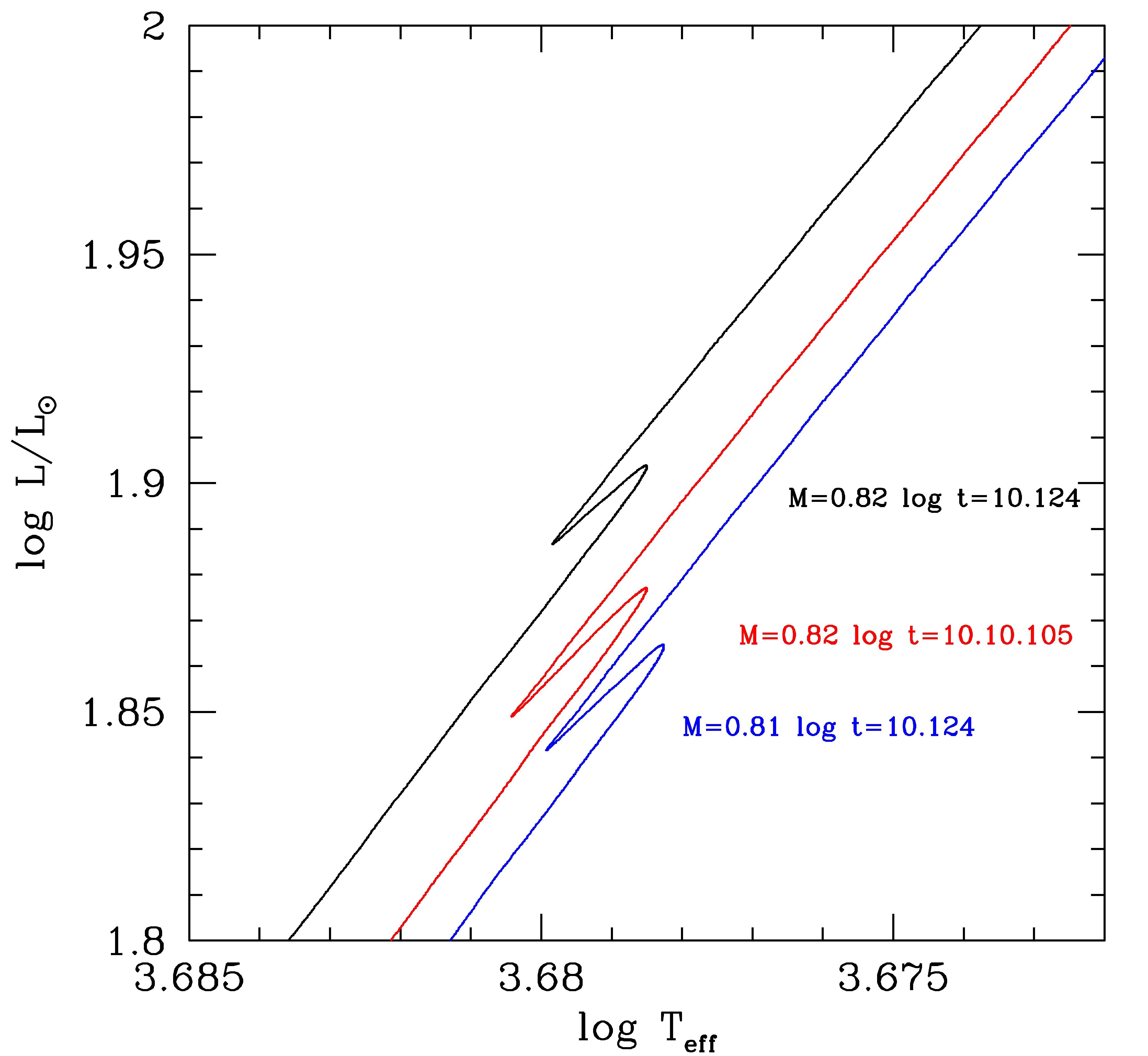}
    \vspace{2mm} 
    \includegraphics[width=0.8\columnwidth]{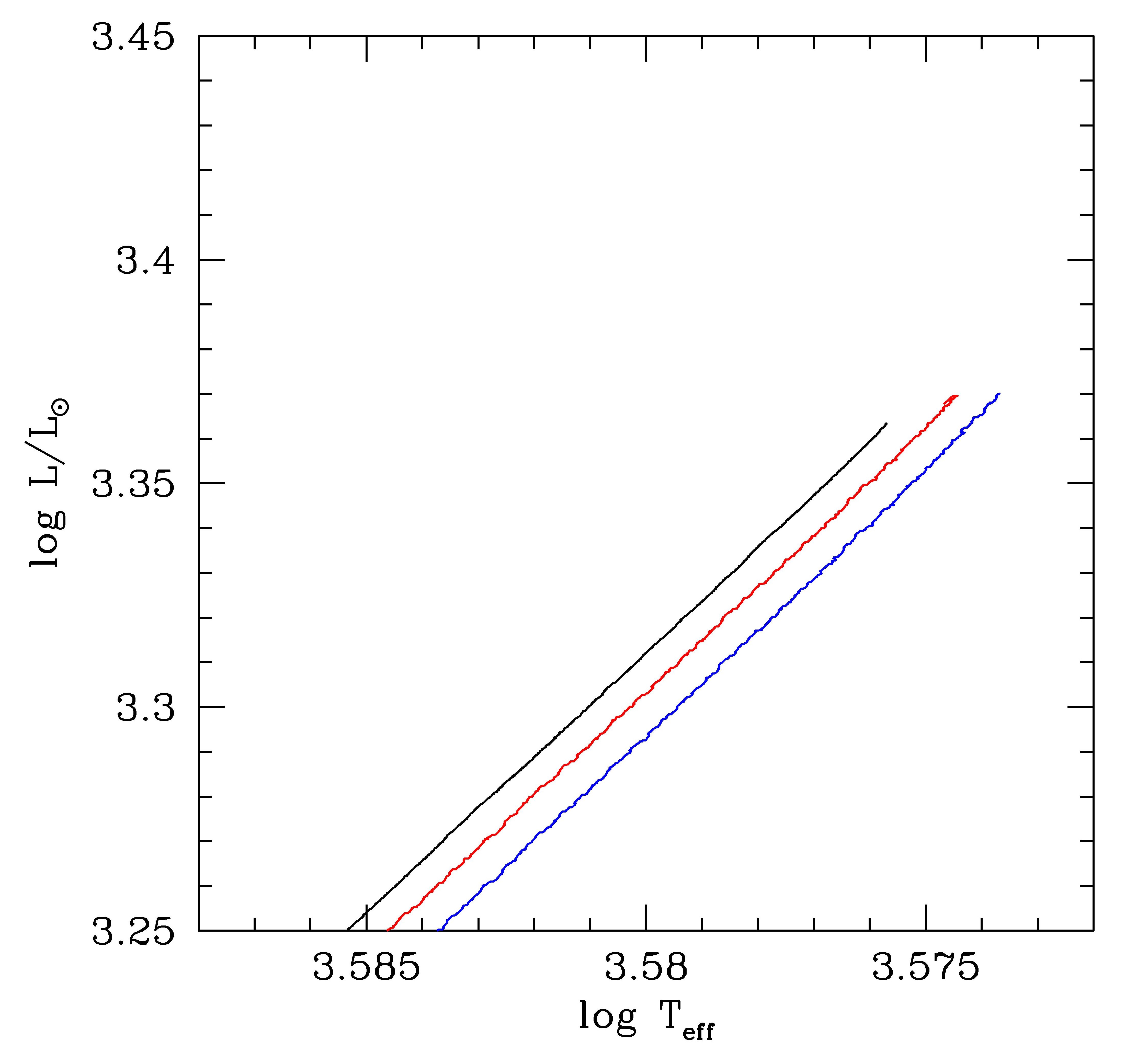}
    
    \caption{Comparisons among evolutionary tracks with and without microscopic diffusion. The upper and the lower panels show the RGBB and TRGB regions of the $\log L - \log T_{eff}$ diagram, respectively. The black line represents the model without microscopic diffusion, while blue and red lines indicate models including microscopic diffusion. Note that $\log T_{eff}$ offsets of -0.0008 and -0.0012 have been applied to the red and the blue evolutionary tracks, respectively. }
    \label{fig:diff_conf}
\end{figure}

We use the FuNS (Full Network Stellar evolution) code to calculate evolutionary tracks of stars whose masses (or ages) fall within the range of GC RGB stars. The FuNS version here adopted is the same as in \cite{straniero_2020} with a few exceptions. In particular, a finer grid of the thermodynamic tables used to calculate the density, the specific heats and the adiabatic temperature gradient, as a function of P, T and chemical composition, is adopted. In addition, we use a cubic  spline interpolation in both T and P, instead of the linear one adopted in \cite{straniero_2020}. Moreover, according to recent R-matrix studies by \cite{Skowronski_2023PhRvL} and \cite{rapagnani_2025PhRvC}, some nuclear reaction rates of the CNO cycle have been updated, specifically: $^{12}$C$(p,\gamma)^{13}$N , $^{13}$C$(p,\gamma)^{14}$N, $^{17}$O$(p,\gamma)^{18}$F and $^{17}$O$(p,\alpha)^{14}$N. The other nuclear reaction rates of the pp chain and the CNO cycle, as well as that of the $3\alpha$ reaction, are from the STARLIB repository. The entire nuclear network includes 33 isotopes, from H up to $^{28}\text{Si}$.
Although these changes have a minor impact on the predicted RGBB and TRGB luminosities, they contribute to reducing the overall theoretical uncertainty. The theoretical predictions, however, are more sensitive to the adopted temporal and spatial resolution of the stellar models. For the evolutionary tracks here presented, the whole RGB sequence is covered with about $2\times10^4$ stellar models, $\sim1.5\times10^4$ from the bump to the tip. The spatial resolution is controlled by an adaptive mesh-point algorithm tuned to increase the resolution around the inner border of the convective envelope and in the shell-burning zone. The number of mesh-points ranges from 900 to 1100.
 Models with and without microscopic diffusion have been computed. This process, as driven by pressure, temperature and chemical gradients, modifies the internal chemical stratification during the long main-sequence phase, and its effects are only partially canceled when the external convection penetrates inward at the first dredge up epoch.  The diffusion coefficients are computed following the prescriptions of \cite{Thoul_1994ApJ}. While this phenomenon marginally affects TRGB predictions, it leads to a non-negligible reduction in RGBB luminosity. This occurrence is illustrated in Fig. \ref{fig:diff_conf}. For a fixed age, models with microscopic diffusion predict a fainter RGBB ($\delta M_{RGBB}\sim 0.1$ mag) and a brighter TRGB ($\delta M_{TRGB}\sim 0.017$ mag). As suggested by helioseismic studies \citep{christensen_1993ApJ, bahcall_1995RvMP}, inclusion of diffusion results in a significant improvement in the agreement between theory and observations. Since the Sun is a low-mass star whose structure is similar to that of a main-sequence GC star, we adopt models including microscopic diffusion as our fiducial choice. These models also provide a better match to the observed RGBB luminosity, further supporting the inclusion of diffusion in our analysis (see Sect. \ref{sect:RGBB_reliability}).
 
We run evolutionary tracks from the pre-main-sequence (first hydrostatic fully convective model) to the TRGB, for three different metallicities, specifically: $Z=1.3\times10^{-3}$, $2\times10^{-3}$ and $5\times10^{-3}$. The initial mass ranges are designed to cover RGB ages between 10 and 13.5 Gyr. For the majority of the evolutionary tracks, the initial He abundance is $Y=0.25$, close to the latest primordial He abundance determinations \citep[see][and reference therein]{Aver_2026arXiv}. A few additional tracks are computed for an enhanced He abundance (up to $Y=0.27$). The RGBB is brighter in He-enhanced tracks of the same age, specifically showing $\delta M_{RGBB}^{25-27}\sim 0.03$. The tip, on the other hand, is less bright by $\delta M_{TRGB}^{25-27}\sim -0.01$, so $\Delta M=M_{TRGB}-M_{RGBB}$ increases.
In general, a solar-calibrated mixing length parameter is adopted, specifically $\alpha_{ML}=1.82$. Nevertheless, in order to check the influence of this choice on our results, we have varied $\alpha_{ML}$ from 1.6 to 1.94. As a result, we find that the RGBB is slightly fainter when $\alpha_{ML}$ is reduced ($\delta M_{RGBB}^{1.6-1.94}\sim 0.011$ ), while the effect on the TRGB is negligible.  

\begin{figure}[htb]
    \centering
    \includegraphics[width=0.9\columnwidth]{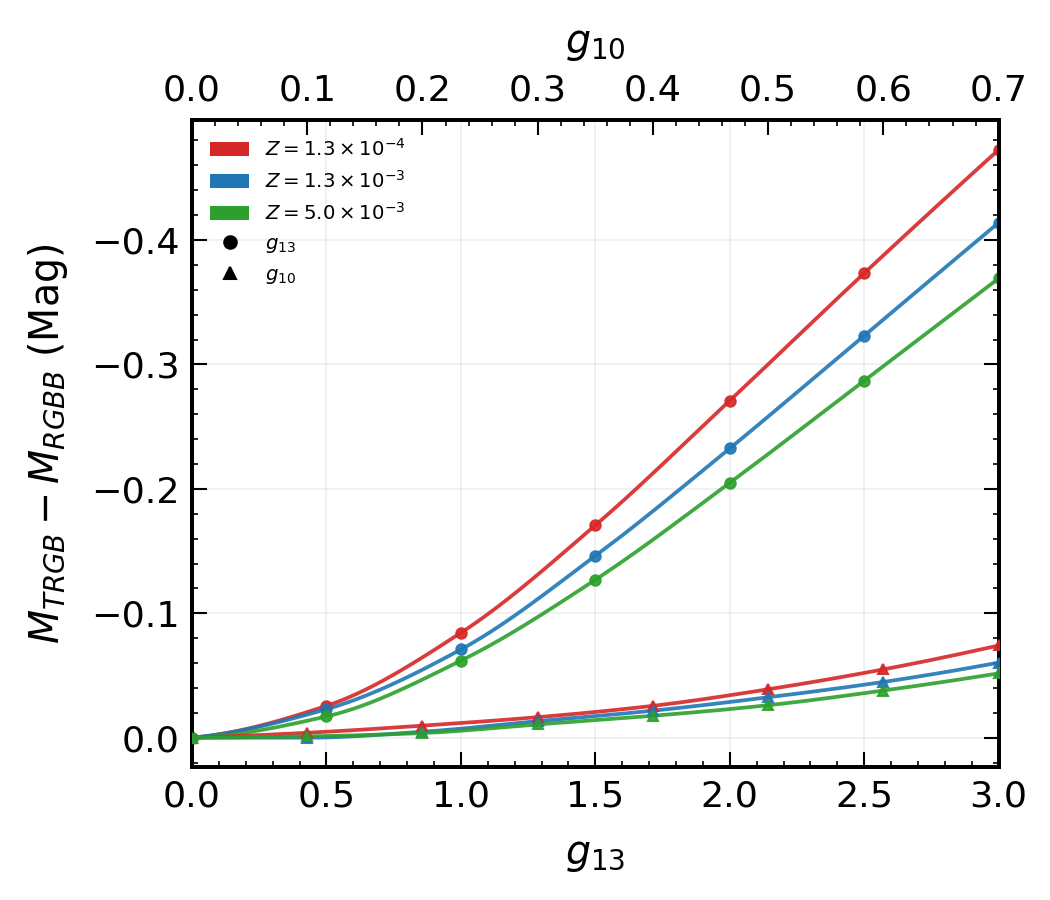}
    \caption{$\Delta M_{TRGB-RGBB}$ with offset to the standard physics value ($g_{10}=g_{13}=0$) for models with $Y=0.25$ and an age of roughly 13.5 Gyr. Both coupling are switched on independently. We see that close to the current best bounds for each coupling, the TRGB is much more sensitive to the axion-electron coupling. In any case, the presence of a small axion-photon coupling would lead to a more stringent constrain on $g_{13}$, thus we are being conservative by fixing $g_{10}=0$.}
    \label{fig:deltaM_g10_13}
\end{figure}

 Axion energy-loss rates for thermal processes possibly active before and during the RGB evolution are computed as in \cite{Straniero_2019}. It includes axions from Primakoff, Compton and Bremsstrahlung processes, the latter being the most important for RGB stars (see the Appendix). Then, we  vary the axion-electron coupling strength $g_{13}=g_{ae}/10^{-13}$ in the range $[0.0, 3.0]$ keeping the axion-photon coupling $g_{10}=g_{a\gamma}/10^{-10} {\rm GeV}=0$. To assess the relevance of a non-negligible photon coupling, we also present a set of evolutionary tracks for $g_{10}$ in the range $[0.0, 0.7]$, which roughly corresponds to the currently unconstrained range of $g_{a\gamma}$, while keeping $g_{13}=0$ (see Fig.~\ref{fig:deltaM_g10_13}). In any case, an additional energy loss due to a non-zero axion-photon coupling would only lead to a tighter bound on the axion-electron coupling. Therefore, setting $g_{a\gamma}=0$ provides a conservative constraint on $g_{ae}$.

\section{Can stellar models reproduce the RGBB brightness?}\label{sect:RGBB_reliability}
A long-standing debate remains open on whether a systematic discrepancy exists between the predicted and observed luminosities of the red giant branch bump.
Soon after the RGBB of GCs were first observed, there has been claims that canonical theoretical stellar models predict a RGBB 0.4 mag too bright \citet{fusipecci_1990}. This result was revised by  \citet{Cassisi-Innocenti-Salaris}, \citet{Cassisi-Salaris} and \cite{ferraro_1999AJ}, who found a good agreement between theoretical prediction and observation once all uncertainties and the global metallicity of individual clusters are taken into account. \citet{Riello}, in particular, found a good agreement for clusters with $\text{[M/H]}\geq-1.75\text{dex}$. However, later studies including the metal poor clusters \cite{Cassisi2011}, \cite{2010ApJ...712..527D}, and \cite{Meissner2006} found a discrepancy of $\sim 0.2\,\text{mag}$ between prediction and observation.  Moreover, some authors noticed that models predict fainter than observed RGBB for a few very metal-rich GCs \citep[see][and references therein]{Joyce_Chaboyer}, while  \cite{nataf_2013ApJ} argue for a second-parameter problem. Summarizing, for metal-rich clusters, i.e., $\text{[M/H]} > -1.5$, theory and observations appear to be in better agreement, while a clear discrepancy is found for both the metal-poor (over bright predictions) and the most metal-rich (under bright predictions) ones.

In general, the RGBB luminosity depends on the cluster age and the original chemical composition ([Fe/H], [$\alpha$/Fe], and helium content). In addition, the observed brightness should be properly rescaled, taking into account the cluster distance and the extinction. All these quantities are obviously affected by uncertainties that can lead to systematic errors in the determination of the RGBB brightness. Additionally, as shown in Section~\ref{sect:models}, the predicted RGBB luminosity depends on the efficiency of microscopic diffusion during the long main-sequence phase, which alters the chemical profile below the convective envelope and leads to a fainter RGBB. Moreover, \citet{Alongi} showed that the activation of overshooting at the inner border of the convective envelope may eventually resolve the discrepancy for those clusters whose RGBB appears fainter than predicted. However, it seems difficult to justify cluster-to-cluster variations in the extent and/or efficiency of this convective overshooting.    

It should be also recalled that these studies usually compare the observed RGBB magnitude in the V (or I) band with the one predicted by models. As we have already stressed, this procedure requires the effective temperature to calculate the bolometric correction; thus, the result depends on many stellar model ingredients, such as the adopted $\alpha_{\text{ML}}$ parameter (mixing length), low-temperature molecular and atomic opacity, and the outer boundary conditions imposed to close the model integration, whose uncertainties may substantially affect the comparison. Instead a comparison based on the luminosity (or $M_{bol}$) prediction is generally more robust. 
 Since we intend to use the RGBB as a reference point to study beyond-the-standard-model physics, we aim to test whether our models provide RGBB bolometric magnitude predictions that match observations or, instead, we are introducing systematic uncertainties in our analysis.

\begin{figure}[htb]
    \centering
    \includegraphics[width=0.8\columnwidth]{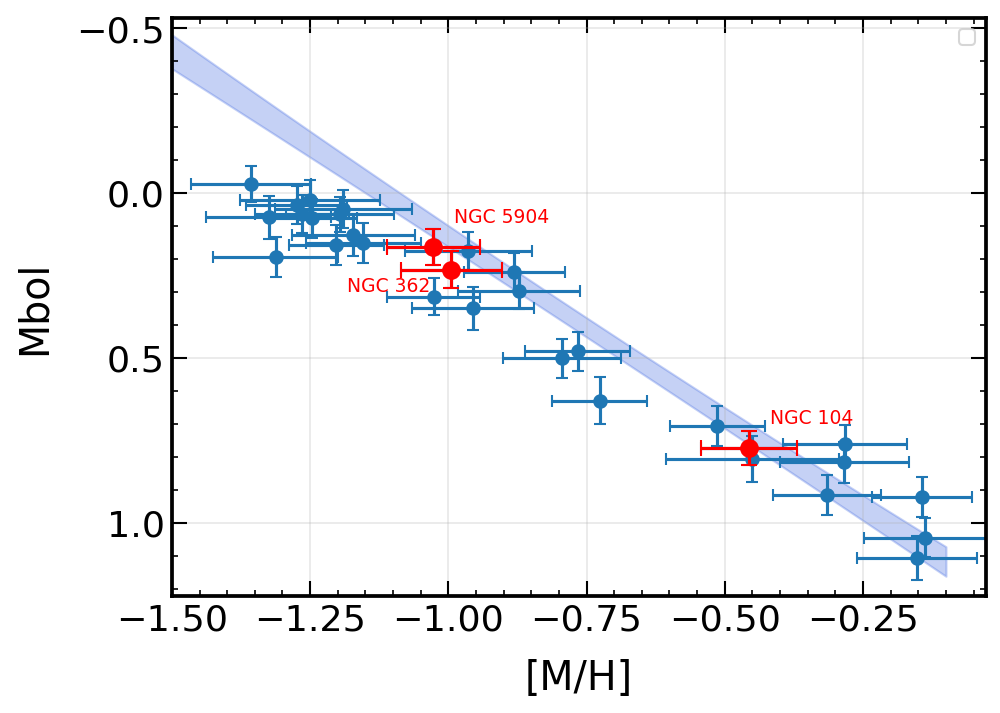}
    \includegraphics[width=0.8\columnwidth]{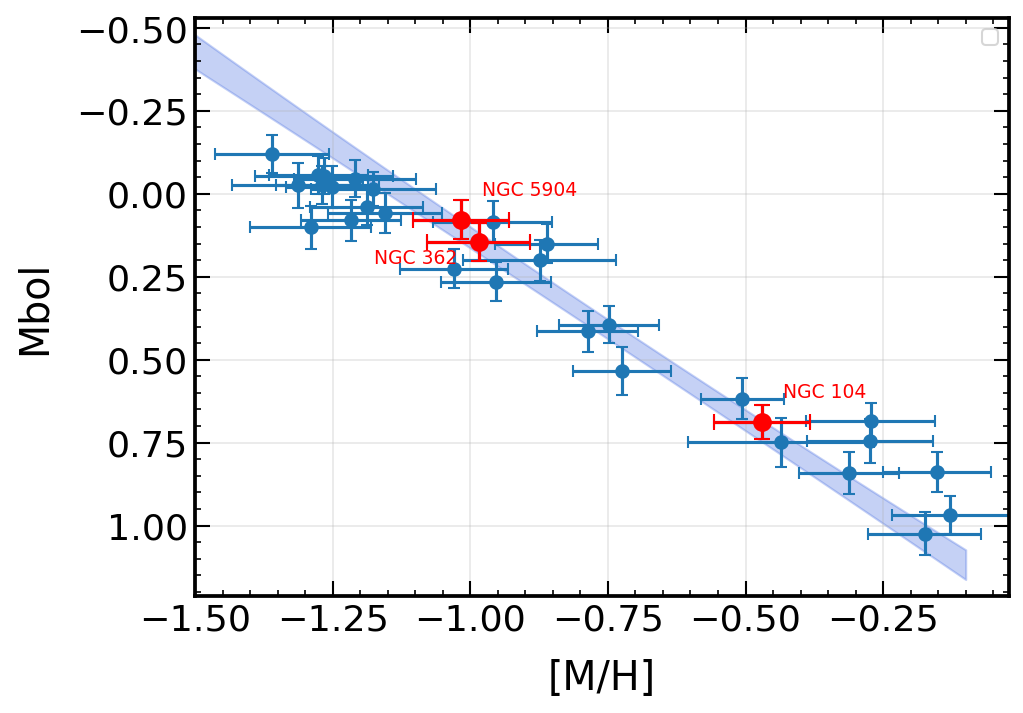}
    \includegraphics[width=0.8\columnwidth]{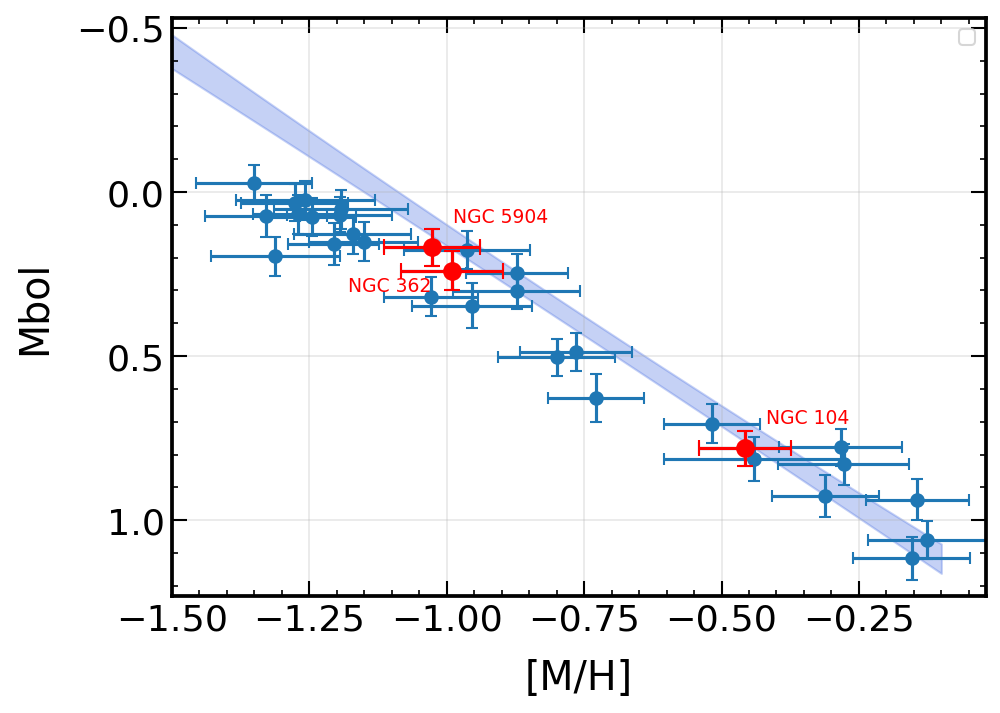}
    
    \caption{The observed RGBB bolometric magnitudes of 31 GCs, as derived from the sample reported by \cite{nataf_2013ApJ}, are compared to our theoretical predictions for ages between 11 and 13 Gyr and $Y=0.25$ (shaded area). Models includes effects of  microscopic diffusion and.  Bolometric corrections are from \cite{worthey_2011} (upper panel) and \cite{castelli_2003} (middle panel) and \citep{MARCS} (bottom panel).
    }
    \label{fig:m_rgbb}
\end{figure}

 We start from the list of RGBB $V$ apparent magnitudes and $V-I$ colors provided by \citet{nataf_2013ApJ}, which contains data for 72 GCs. We then limit our analysis to GCs with a metallicity in the same range as NGC 362, NGC 5904, and NGC 104. Hence, coherently with our approach in Sect.~\ref{sec:obsv_RGBB}, distance moduli are derived from \citet{baumgardt_2021MNRAS}, $\text{[Fe/H]}$ from \citet{carretta_2009AA}, and reddening from \citet{Harris_2010}. Moreover, for all clusters, we assume $\alpha/\mathrm{Fe}]=0.4$. With these inputs, we compute bolometric corrections and, in turn, the corresponding $M_{\text{bol}}$ for each cluster, according to \citet{worthey_2011} and \citet{castelli_2003}.
 The comparisons between the theoretical predictions (for ages ranging from 11 to 13 Gyr, and $Y=0.25$) and the observed RGBB luminosities (for both sets of bolometric corrections) are shown in Fig.~\ref{fig:m_rgbb}. We find that our predictions match the observations quite well for $\text{[M/H]} > -1.5$, but we also confirm previous findings that the predicted RGBB magnitudes are too bright at lower metallicities.  As previously mentioned, this is the reason why we decided to restrict our study to this metallicity range. We also find that the theoretical bolometric corrections by \citet{castelli_2003} yield a better agreement between predictions and observations. In particular, the RGBB luminosities of the three clusters used in the present work to constrain the axion-electron coupling are in agreement with our theoretical predictions, independently of the adopted bolometric correction set.

\section{Statistical analysis and results}\label{sect:statistics}

With the TRGB and RGBB measurements available for the three clusters, we now compare our theoretical models, with and without axions, to the observations. We perform a standard maximum likelihood analysis, with uncertainty propagation carried out through a Monte Carlo approach.

\begin{figure}
    \centering
    \includegraphics[width=\linewidth]{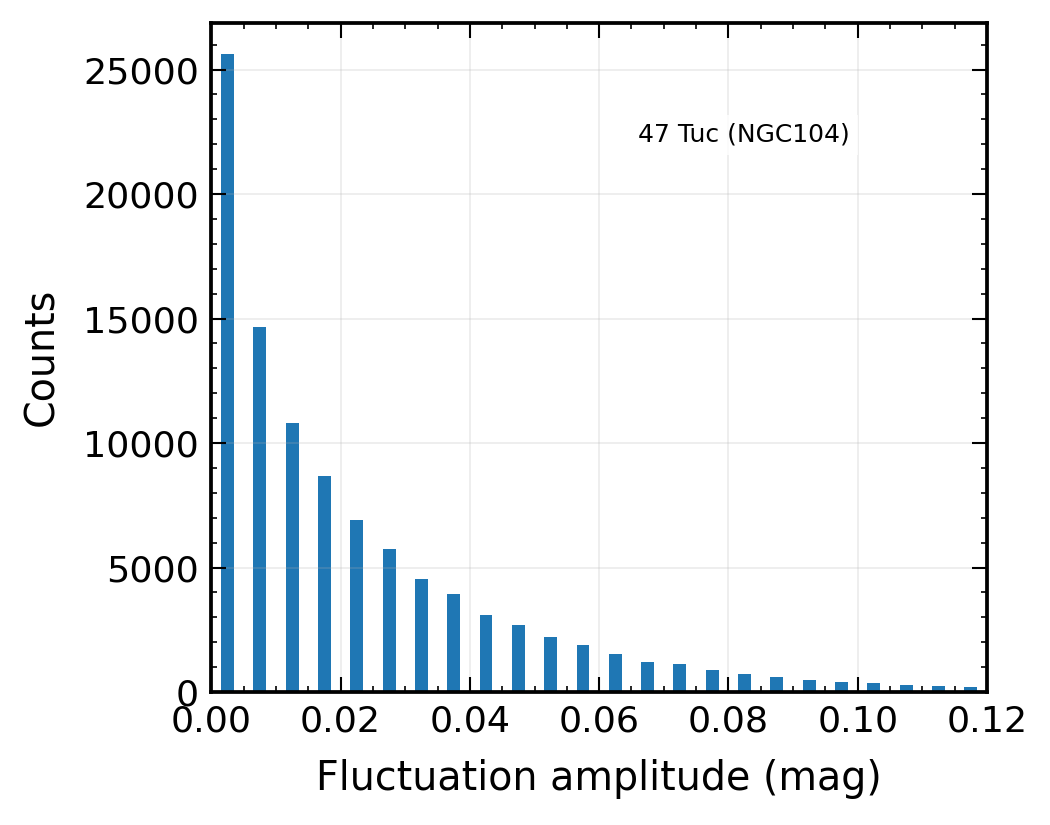}
    \caption{TRGB correction for 47TUC}
    \label{fig:TRGB correction}
\end{figure}

For each cluster and each value of $g_{13}$, we perform 10,000 simulations in which [M/H], the age, $E(B-V)$, and the $V$- and $I$-band magnitudes of the TRGB and RGBB are drawn from Gaussian distributions centered on their measured values, with widths given by the corresponding uncertainties (see Tables \ref{tab:clusterparameters}--\ref{tab:tipmag}). The helium mass fraction $Y$ is drawn from a uniform distribution in the range 0.2450--0.2550 and the GC's age from a uniform distribution in the range 11 Gyr--13 Gyr. The statistical correction to the TRGB is taken from the corresponding distribution obtained for each cluster (see for e.g Fig.\ref{fig:TRGB correction}).

\begin{figure}
    \centering
    \includegraphics[width=1\columnwidth]{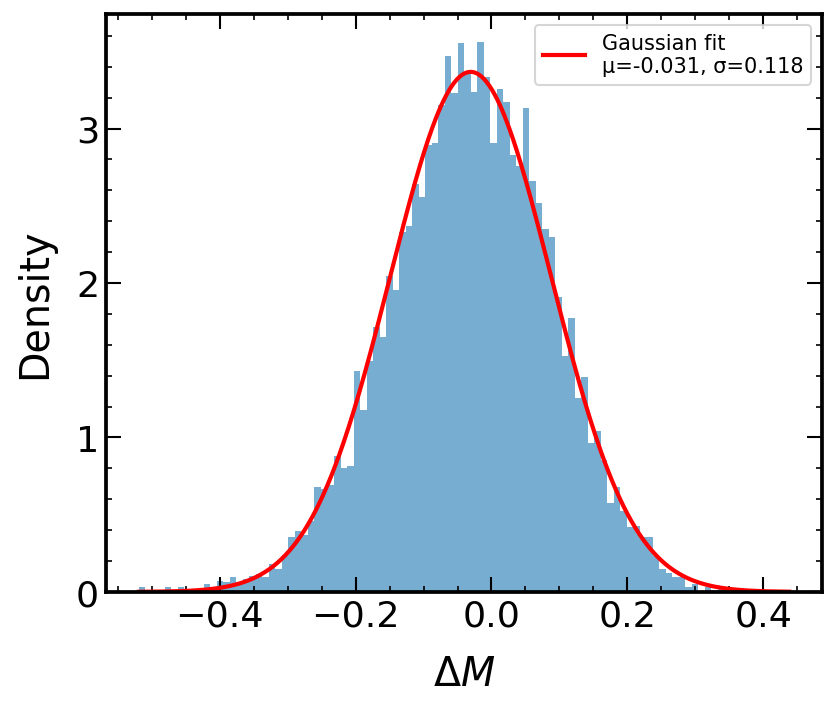}
    
    \caption{Distribution of the residue $\delta(g_{13})$, here in the case $g_{13}=0$ for 47tuc. One can see that it is well approximated by  a gaussian described by the two parameter $\mu(g_{13})$ and $\sigma(g_{13})$
    }
    \label{fig:gaussian_fit}
\end{figure}

For each set of parameters and each value of $g_{13}$, we compute the residual
\[
 \Delta_i(g_{13}) = \Delta M^{\mathrm{obs}}_i - \Delta M^{\mathrm{th}}_i(g_{13}),
\]

where $\Delta M = M_{\mathrm{TRGB}} - M_{\mathrm{RGBB}}$. 
For each $g_{13}$, the distribution of $\Delta_i(g_{13})$ is very well approximated by a Gaussian, see for e.g Fig.\ref{fig:gaussian_fit}. We therefore characterize it by its mean $\mu(g_{13})$ and standard deviation $\sigma(g_{13})$, both extracted from the same Monte Carlo sample. The likelihood for $g_{13}$ is then written as:
\begin{equation}
    \mathcal{L}(g_{13}) =
    \frac{1}{\sqrt{2\pi}\,\sigma(g_{13})}
    \exp\left[
    -\frac{\mu(g_{13})^2}{2\sigma^2(g_{13})}
    \right].
\end{equation}
We adopt a flat prior on $g_{13}$ over the range $0\leq g_{13}\leq 3$. The $95\%$ upper limit is then defined as the value $g_{95}$ satisfying:
\begin{equation}
    \frac{\int_0^{g_{95}} \mathcal{L}(g_{13}),\mathrm{d}g_{13}}
{\int_0^3 \mathcal{L}(g_{13}),\mathrm{d}g_{13}}=0.95
\end{equation}

\begin{figure}
    \centering
    \includegraphics[width=\columnwidth]{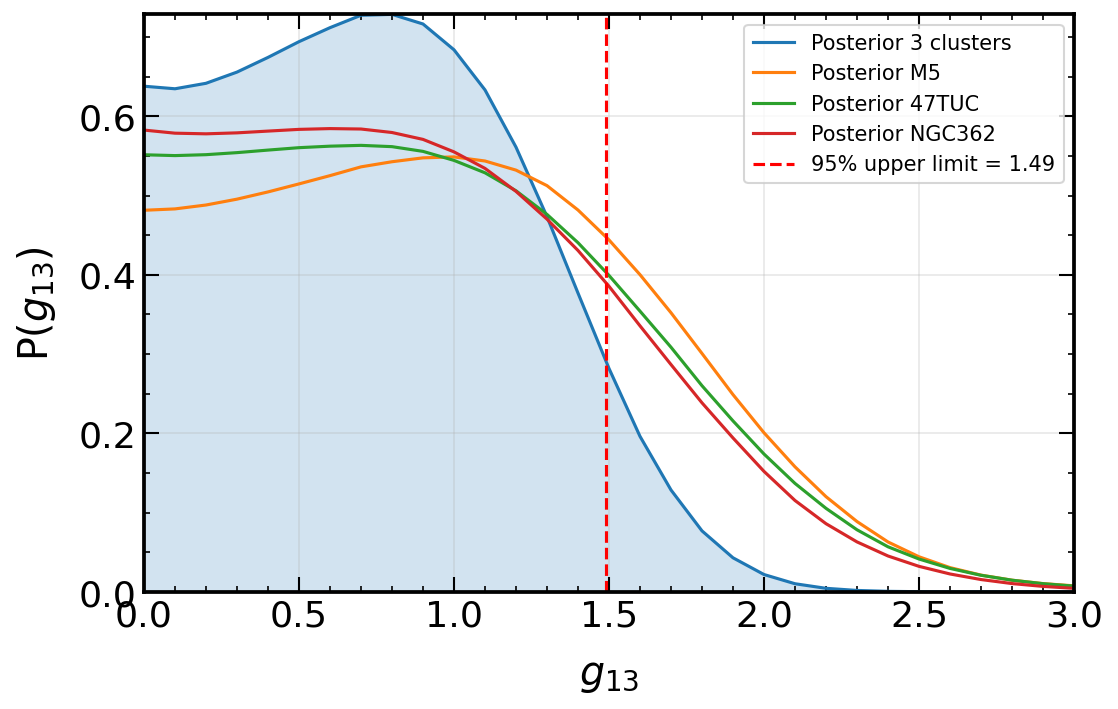}
    
    \caption{Likelihood function for 47tuc, NGC362 and M5 as well as joint likelihhod for the \citet{worthey_2011} BCs. The 95\% confidence level bound is also shown.
    }
    \label{fig:likelihood_WL2011}
\end{figure}

\begin{figure}
    \centering
    \includegraphics[width=\columnwidth]{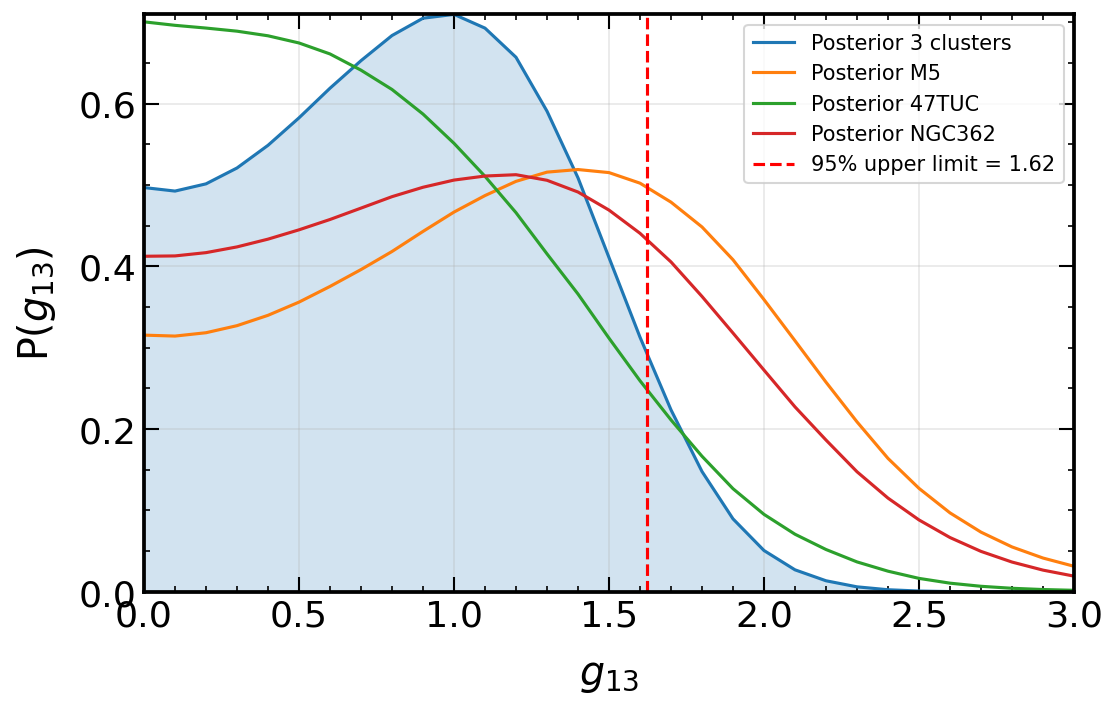}
    
    \caption{Likelihood function for 47tuc, NGC362 and M5 as well as joint likelihhod for the MARCS BCs \citep{MARCS}. The 95\% confidence level bound is also shown.
    }
    \label{fig:likelihood_MARCS}
\end{figure}

This corresponds to a Bayesian $95\%$ upper credible bound with a prior uniform in $g_{13}$. We perform the analysis independently for the two sets of BCs, i.e \cite{worthey_2011} and \cite{MARCS}. With this prescription, the maximum-likelihood values are $g_{13}=0.7$, 0.6, and 1.0 for 47 Tucanae, NGC 362, and M5, respectively, while the corresponding $95\%$ upper limits are 1.99, 1.92, and 2.03 for \cite{worthey_2011}. The joint likelihood analysis yields a maximum likelihood at $g_{13} = 0.8$ and a combined 95\% upper bound of 1.49 (Fig.\ref{fig:likelihood_WL2011}). For \cite{MARCS},  the maximum-likelihood values are $g_{13}=0.0$, 1.2, and 1.4 for 47 Tucanae, NGC 362, and M5, respectively, while the corresponding $95\%$ upper limits are 1.76, 2.23, and 2.36. The joint likelihood analysis yields a maximum likelihood at $g_{13} = 0.9$ and a combined 95\% upper bound of 1.62  (Fig.\ref{fig:likelihood_MARCS}).

\section{Summary and Conclusions}\label{sec:conclusions}

In this paper, we have presented a novel and alternative approach to constrain the axion-electron coupling constant ($g_{ae}$) using RGB stars in Galactic globular clusters. The primary goal of this work was to mitigate the dominant sources of systematic uncertainties, most notably distance scale dependencies and the zero-point of bolometric corrections, that may have limited the precision of previous observational bounds.  Rather than relying on the absolute luminosity of the tip of the red giant branch (TRGB) alone, our methodology exploits a distance-independent observable: the bolometric magnitude difference between the RGB bump and the TRGB, denoted as $\Delta M = M_{\text{TRGB}} - M_{\text{RGBB}}$. While an anomalous energy loss via thermal axion bremsstrahlung forces the degenerate helium core to grow more massive before helium ignition making the TRGB brighter, the RGB bump luminosity remains unaffected by a non-zero axion-electron coupling. Consequently, any variation in $\Delta M$ can be directly attributed to the extra cooling channel induced by axions. 

To test the robustness of our theoretical framework, we used the FuNS code to investigate the potential discrepancies between predicted and observed RGBB luminosities. Accounting for the fundamental role of microscopic diffusion (which, for a given age, causes a fainter RGBB by $\sim 0.1$ mag), we confirmed that our stellar models accurately reproduce the observed RGBB brightness for relatively metal-rich clusters ($[M/\mathrm{H}] > -1.5$). For this reason, we restricted our statistical analysis to three specific intermediate- and high-metallicity clusters: NGC 362, NGC 5904 (M5), and NGC 104 (47 Tuc).  In the models used for our analysis, the inner boundary of the convective envelope is determined using the classical Schwarzwild criterion, and no overshooting was applied. The effect of the latter phenomenon would be to reduce the bump luminosity and, consequently, to increase the theoretical $\Delta M$. In practice, a non-zero overshooting would lead to a reduction of the difference with the observed $\Delta M$ values, thereby yielding a more stringent bound. It goes without saying that the bound obtained with models that do not include any overshoot is the most conservative one. On the other hand, it should be noted that the extent of any potential overshoot, as well as the efficiency of mixing and energy transport in this region, cannot be determined from first principles; therefore, any specific choice remains arbitrary.

A rigorous maximum likelihood analysis, combined with Monte Carlo simulations to handle uncertainty propagation, yields a global maximum likelihood at $g_{13} = 0.8$ with a combined 95\% C.L. upper limit of $1.49$ when the WL2011 bolometric corrections are adopted, and $g_{13} = 0.9.$ (1.62 at 95\% C.L.) when MARCS BCs are considered. The difference in the results provides a direct indication of the magnitude of the systematic uncertainty associated to the bolometric correction, which has not always been appropriately considered in previous studies.



While this new bound is slightly less stringent than some recent global analyses using large multi-cluster samples, it represents a step forward in terms of reliability. Because our method bypasses the distance scale entirely and reduces other systematic effects such as those due to the light extinction and the calibration of the bolometric correction (BC zero point), this limit stands as one of the most robust constraints on the axion-electron coupling derived from stellar astrophysics to date. By applying this differential technique to a larger sample of clusters we hope to increase the statistical significance of the analysis and reduce cluster-specific statistical uncertainties. However, the improvement will ultimately be limited by the cumulative systematic uncertainties shared across clusters, including those associated with the input physics of stellar models, bolometric corrections, the age and abundance scales. Given the current level of observational and theoretical uncertainties, corresponding to an uncertainty of order 0.1 mag in the TRGB magnitude, we do not expect to be able to set substantially tighter constraints beyond $g_{13}\sim1.0$ unless these systematic uncertainties can be reduced.

We notice that very recently, \citet{Gontcharov:2026knx} analyzed the TRGB luminosities of 27 Galactic globular clusters using cluster-specific isochrone fits and dedicated MESA calculations, reporting the stringent upper limit ($g_{ae}<3.8\times10^{-14}$) at $95\%$ confidence. This result should, however, be interpreted with caution. In their joint likelihood, the observational and theoretical uncertainties are included separately in the variance assigned to each cluster, and the individual cluster likelihoods are subsequently multiplied. Several important contributions to the theoretical uncertainty budget, including those associated with neutrino emission rates, conductive opacities, nuclear reaction rates, screening prescriptions, and the equation of state, originate from common stellar-physics inputs and are therefore correlated across the cluster sample. Although the corresponding shift in the TRGB luminosity may vary with the stellar parameters of each cluster, the underlying nuisance parameters are shared. Treating these uncertainties as statistically independent effectively allows common theoretical systematics to average down when the clusters are combined, potentially producing an artificially narrow joint likelihood and an overly stringent upper limit. A consistent multi-cluster analysis should instead introduce the common stellar-physics inputs as shared nuisance parameters, or equivalently include their correlations through a non-diagonal covariance matrix.

\begin{acknowledgements}
This article is based upon work from COST Action COSMIC WISPers CA21106, supported by COST (European Cooperation in Science and Technology). AC is supported by an ERC STG grant (“AstroDarkLS”, grant No. 101117510). AC also acknowledges the support by the European Research Area (ERA) via the UNDARK project (project number 101159929). AC and TL also acknowledge the Weizmann Institute of Science for hospitality in the final stages of this project.
\end{acknowledgements} 
   \bibliographystyle{aa} 
   \bibliography{Biblio.bib} 

@ARTICLE{tailo2020,
       author = {{Tailo}, M. and {Milone}, A.~P. and {Lagioia}, E.~P. and {D'Antona}, F. and {Marino}, A.~F. and {Vesperini}, E. and {Caloi}, V. and {Ventura}, P. and {Dondoglio}, E. and {Cordoni}, G.},
        title = "{Mass-loss along the red giant branch in 46 globular clusters and their multiple populations}",
      journal = {\mnras},
         year = 2020,
        month = nov,
       volume = {498},
       number = {4},
        pages = {5745-5771},
          doi = {10.1093/mnras/staa2639},
archivePrefix = {arXiv},
       eprint = {2009.01080},
 primaryClass = {astro-ph.SR},
       adsurl = {https://ui.adsabs.harvard.edu/abs/2020MNRAS.498.5745T}
}

@ARTICLE{mcdonald2015,
       author = {{McDonald}, I. and {Zijlstra}, A.~A.},
        title = "{Mass-loss on the red giant branch: the value and metallicity dependence of Reimers' {\ensuremath{\eta}} in globular clusters}",
      journal = {\mnras},
         year = 2015,
        month = mar,
       volume = {448},
       number = {1},
        pages = {502-521},
          doi = {10.1093/mnras/stv007},
archivePrefix = {arXiv},
       eprint = {1501.00874},
 primaryClass = {astro-ph.SR},
       adsurl = {https://ui.adsabs.harvard.edu/abs/2015MNRAS.448..502M}
}

@ARTICLE{christensen_1993ApJ,
       author = {{Christensen-Dalsgaard}, J. and {Proffitt}, C.~R. and {Thompson}, M.~J.},
        title = "{Effects of Diffusion on Solar Models and Their Oscillation Frequencies}",
      journal = {\apjl},
         year = 1993,
        month = feb,
       volume = {403},
        pages = {L75},
          doi = {10.1086/186725},
       adsurl = {https://ui.adsabs.harvard.edu/abs/1993ApJ...403L..75C}
}

@ARTICLE{bahcall_1995RvMP,
       author = {{Bahcall}, John N. and {Pinsonneault}, M.~H. and {Wasserburg}, G.~J.},
        title = "{Solar models with helium and heavy-element diffusion}",
      journal = {Reviews of Modern Physics},
         year = 1995,
        month = oct,
       volume = {67},
       number = {4},
        pages = {781-808},
          doi = {10.1103/RevModPhys.67.781},
archivePrefix = {arXiv},
       eprint = {hep-ph/9505425},
 primaryClass = {hep-ph},
       adsurl = {https://ui.adsabs.harvard.edu/abs/1995RvMP...67..781B}
}

@ARTICLE{ferraro_1999AJ,
       author = {{Ferraro}, F.~R. and {Messineo}, M. and {Fusi Pecci}, F. and {de Palo}, M.~A. and {Straniero}, O. and {Chieffi}, A. and {Limongi}, M.},
        title = "{The Giant, Horizontal, and Asymptotic Branches of Galactic Globular Clusters. I. The Catalog, Photometric Observables, and Features}",
      journal = {\aj},
         year = 1999,
        month = oct,
       volume = {118},
       number = {4},
        pages = {1738-1758},
          doi = {10.1086/301029},
archivePrefix = {arXiv},
       eprint = {astro-ph/9906248},
 primaryClass = {astro-ph},
       adsurl = {https://ui.adsabs.harvard.edu/abs/1999AJ....118.1738F}
}

@article{Irastorza:2021tdu,
    author = "Irastorza, Igor Garcia",
    title = "{An introduction to axions and their detection}",
    eprint = "2109.07376",
    archivePrefix = "arXiv",
    primaryClass = "hep-ph",
    doi = "10.21468/SciPostPhysLectNotes.45",
    journal = "SciPost Phys. Lect. Notes",
    volume = "45",
    pages = "1",
    year = "2022"
}

@article{DiLuzio:2020wdo,
    author = "Di Luzio, Luca and Giannotti, Maurizio and Nardi, Enrico and Visinelli, Luca",
    title = "{The landscape of QCD axion models}",
    eprint = "2003.01100",
    archivePrefix = "arXiv",
    primaryClass = "hep-ph",
    reportNumber = "DESY 20-036, DESY-20-036",
    doi = "10.1016/j.physrep.2020.06.002",
    journal = "Phys. Rept.",
    volume = "870",
    pages = "1--117",
    year = "2020"
}

@article{OHare:2024nmr,
    author = "O'Hare, Ciaran A. J.",
    title = "{Cosmology of axion dark matter}",
    eprint = "2403.17697",
    archivePrefix = "arXiv",
    primaryClass = "hep-ph",
    doi = "10.22323/1.454.0040",
    journal = "PoS",
    volume = "COSMICWISPers",
    pages = "040",
    year = "2024"
}

@ARTICLE{Skowronski_2023PhRvL,
       author = {{Skowronski}, J. and {Boeltzig}, A. and {Ciani}, G.~F. and {Csedreki}, L. and {Piatti}, D. and {Aliotta}, M. and {Ananna}, C. and {Barile}, F. and {Bemmerer}, D. and {Best}, A. and {Broggini}, C. and {Bruno}, C.~G. and {Caciolli}, A. and {Campostrini}, M. and {Cavanna}, F. and {Colombetti}, P. and {Compagnucci}, A. and {Corvisiero}, P. and {Davinson}, T. and {Depalo}, R. and {Di Leva}, A. and {Elekes}, Z. and {Ferraro}, F. and {Formicola}, A. and {F{\"u}l{\"o}p}, Zs. and {Gervino}, G. and {Gesu{\`e}}, R.~M. and {Guglielmetti}, A. and {Gustavino}, C. and {Gy{\"u}rky}, Gy. and {Imbriani}, G. and {Junker}, M. and {Lugaro}, M. and {Marigo}, P. and {Masha}, E. and {Menegazzo}, R. and {Paticchio}, V. and {Perrino}, R. and {Prati}, P. and {Rapagnani}, D. and {Rigato}, V. and {Schiavulli}, L. and {Sidhu}, R.~S. and {Straniero}, O. and {Sz{\"u}cs}, T. and {Zavatarelli}, S. and {LUNA Collaboration}},
        title = "{Proton-Capture Rates on Carbon Isotopes and Their Impact on the Astrophysical $^{12}$C / $^{13}$C Ratio}",
      journal = {\prl},
         year = 2023,
        month = oct,
       volume = {131},
       number = {16},
          eid = {162701},
        pages = {162701},
          doi = {10.1103/PhysRevLett.131.162701},
archivePrefix = {arXiv},
       eprint = {2308.16098},
 primaryClass = {nucl-ex},
       adsurl = {https://ui.adsabs.harvard.edu/abs/2023PhRvL.131p2701S}
}

@ARTICLE{Thoul_1994ApJ,
       author = {{Thoul}, Anne A. and {Bahcall}, John N. and {Loeb}, Abraham},
        title = "{Element Diffusion in the Solar Interior}",
      journal = {\apj},
         year = 1994,
        month = feb,
       volume = {421},
        pages = {828},
          doi = {10.1086/173695},
archivePrefix = {arXiv},
       eprint = {astro-ph/9304005},
 primaryClass = {astro-ph},
       adsurl = {https://ui.adsabs.harvard.edu/abs/1994ApJ...421..828T}
}

@ARTICLE{worthey_2011,
       author = {{Worthey}, Guy and {Lee}, Hyun-chul},
        title = "{An Empirical UBV RI JHK Color-Temperature Calibration for Stars}",
      journal = {\apjs},
         year = 2011,
        month = mar,
       volume = {193},
       number = {1},
          eid = {1},
        pages = {1},
          doi = {10.1088/0067-0049/193/1/1},
archivePrefix = {arXiv},
       eprint = {astro-ph/0604590},
 primaryClass = {astro-ph},
       adsurl = {https://ui.adsabs.harvard.edu/abs/2011ApJS..193....1W}
}

@ARTICLE{Harris_2010,
       author = {{Harris}, William E.},
        title = "{A New Catalog of Globular Clusters in the Milky Way}",
      journal = {arXiv e-prints},
         year = 2010,
        month = dec,
          eid = {arXiv:1012.3224},
        pages = {arXiv:1012.3224},
          doi = {10.48550/arXiv.1012.3224},
archivePrefix = {arXiv},
       eprint = {1012.3224},
 primaryClass = {astro-ph.GA},
       adsurl = {https://ui.adsabs.harvard.edu/abs/2010arXiv1012.3224H}
}

@ARTICLE{baumgardt_2021MNRAS,
       author = {{Baumgardt}, H. and {Vasiliev}, E.},
        title = "{Accurate distances to Galactic globular clusters through a combination of Gaia EDR3, HST, and literature data}",
      journal = {\mnras},
         year = 2021,
        month = aug,
       volume = {505},
       number = {4},
        pages = {5957-5977},
          doi = {10.1093/mnras/stab1474},
archivePrefix = {arXiv},
       eprint = {2105.09526},
 primaryClass = {astro-ph.GA},
       adsurl = {https://ui.adsabs.harvard.edu/abs/2021MNRAS.505.5957B}
}

@ARTICLE{carretta_2009AA,
       author = {{Carretta}, E. and {Bragaglia}, A. and {Gratton}, R. and {D'Orazi}, V. and {Lucatello}, S.},
        title = "{Intrinsic iron spread and a new metallicity scale for globular clusters}",
      journal = {\aap},
         year = 2009,
        month = dec,
       volume = {508},
       number = {2},
        pages = {695-706},
          doi = {10.1051/0004-6361/200913003},
archivePrefix = {arXiv},
       eprint = {0910.0675},
 primaryClass = {astro-ph.GA},
       adsurl = {https://ui.adsabs.harvard.edu/abs/2009A&A...508..695C}
}

@ARTICLE{Aver_2026arXiv,
       author = {{Aver}, Erik and {Skillman}, Evan D. and {Pogge}, Richard W. and {Rogers}, Noah S.~J. and {Weller}, Miqaela K. and {Olive}, Keith A. and {Berg}, Danielle A. and {Salzer}, John J. and {Miller}, Jr., John H. and {M{\'e}ndez-Delgado}, Jos{\'e} Eduardo},
        title = "{The LBT Y$_\mathrm{p}$ Project IV: A New Value of the Primordial Helium Abundance}",
      journal = {arXiv e-prints},
         year = 2026,
        month = jan,
          eid = {arXiv:2601.22238},
        pages = {arXiv:2601.22238},
          doi = {10.48550/arXiv.2601.22238},
archivePrefix = {arXiv},
       eprint = {2601.22238},
 primaryClass = {astro-ph.CO},
       adsurl = {https://ui.adsabs.harvard.edu/abs/2026arXiv260122238A}
}

@ARTICLE{rapagnani_2025PhRvC,
       author = {{Rapagnani}, D. and {Straniero}, O. and {Imbriani}, G. and {Aliotta}, M. and {Ananna}, C. and {Barile}, F. and {Barbieri}, L. and {Bemmerer}, D. and {Best}, A. and {Boeltzig}, A. and {Broggini}, C. and {Bruno}, C.~G. and {Caciolli}, A. and {Campostrini}, M. and {Casaburo}, F. and {Cavanna}, F. and {Ciani}, G.~F. and {Colombetti}, P. and {Compagnucci}, A. and {Corvisiero}, P. and {Csedreki}, L. and {Davinson}, T. and {Depalo}, R. and {Di Leva}, A. and {Elekes}, Z. and {Ferraro}, F. and {Formicola}, A. and {F{\"u}l{\"o}p}, Zs. and {Gervino}, G. and {Gesu{\`e}}, R.~M. and {Gy{\"u}rky}, Gy. and {Guglielmetti}, A. and {Gustavino}, C. and {Junker}, M. and {Lugaro}, M. and {Marigo}, P. and {Marsh}, J. and {Masha}, E. and {Menegazzo}, R. and {Mercogliano}, D. and {Paticchio}, V. and {Piatti}, D. and {Prati}, P. and {Rigato}, V. and {Robb}, D. and {Sidhu}, R.~S. and {Skowronski}, J. and {Sz{\"u}cs}, T. and {Zavatarelli}, S. and {LUNA Collaboration}},
        title = "{Revision of the CNO cycle: Rate of O17 destruction in stars}",
      journal = {\prc},
         year = 2025,
        month = feb,
       volume = {111},
       number = {2},
          eid = {025805},
        pages = {025805},
          doi = {10.1103/PhysRevC.111.025805},
       adsurl = {https://ui.adsabs.harvard.edu/abs/2025PhRvC.111b5805R}
}

@ARTICLE{nataf_2013ApJ,
       author = {{Nataf}, David M. and {Gould}, Andrew P. and {Pinsonneault}, Marc H. and {Udalski}, Andrzej},
        title = "{Red Giant Branch Bump Brightness and Number Counts in 72 Galactic Globular Clusters Observed with the Hubble Space Telescope}",
      journal = {\apj},
         year = 2013,
        month = apr,
       volume = {766},
       number = {2},
          eid = {77},
        pages = {77},
          doi = {10.1088/0004-637X/766/2/77},
archivePrefix = {arXiv},
       eprint = {1109.2118},
 primaryClass = {astro-ph.GA},
       adsurl = {https://ui.adsabs.harvard.edu/abs/2013ApJ...766...77N}
}

@INPROCEEDINGS{castelli_2003,
       author = {{Castelli}, F. and {Kurucz}, R.~L.},
        title = "{New Grids of ATLAS9 Model Atmospheres}",
    booktitle = {Modelling of Stellar Atmospheres},
         year = 2003,
       editor = {{Piskunov}, N. and {Weiss}, W.~W. and {Gray}, D.~F.},
       series = {IAU Symposium},
       volume = {210},
        month = jan,
        pages = {A20},
          doi = {10.48550/arXiv.astro-ph/0405087},
archivePrefix = {arXiv},
       eprint = {astro-ph/0405087},
 primaryClass = {astro-ph},
       adsurl = {https://ui.adsabs.harvard.edu/abs/2003IAUS..210P.A20C}
}

@ARTICLE{fusipecci_1990,
       author = {{Fusi Pecci}, F. and {Ferraro}, F.~R. and {Crocker}, D.~A. and {Rood}, R.~T. and {Buonanno}, R.},
        title = "{The variation of the red giant luminosity function ``bump'' with metallicity and the age of the globular clusters.}",
      journal = {\aap},
         year = 1990,
        month = nov,
       volume = {238},
        pages = {95},
       adsurl = {https://ui.adsabs.harvard.edu/abs/1990A&A...238...95F}
}

@inproceedings{Straniero_2018,
    author = "Straniero, Oscar and Dominguez, Inmaculata and Giannotti, Maurizio and Mirizzi, Alessandro",
    title = "{Axion-electron coupling from the RGB tip of Globular Clusters}",
    booktitle = "{13th Patras Workshop on Axions, WIMPs and WISPs}",
    eprint = "1802.10357",
    archivePrefix = "arXiv",
    primaryClass = "astro-ph.SR",
    doi = "10.3204/DESY-PROC-2017-02/straniero_oscar",
    pages = "172--176",
    year = "2018"
}

@ARTICLE{paczynsky_1970,
       author = {{Paczy{\'n}ski}, B.},
        title = "{Evolution of Single Stars. I. Stellar Evolution from Main Sequence to White Dwarf or Carbon Ignition}",
      journal = {\actaa},
         year = 1970,
        month = jan,
       volume = {20},
        pages = {47},
       adsurl = {https://ui.adsabs.harvard.edu/abs/1970AcA....20...47P}
}

@article{Soltis:2020gpl,
    author = "Soltis, John and Casertano, Stefano and Riess, Adam G.",
    title = "{The Parallax of $\omega$ Centauri Measured from Gaia EDR3 and a Direct, Geometric Calibration of the Tip of the Red Giant Branch and the Hubble Constant}",
    eprint = "2012.09196",
    archivePrefix = "arXiv",
    primaryClass = "astro-ph.GA",
    doi = "10.3847/2041-8213/abdbad",
    journal = "Astrophys. J. Lett.",
    volume = "908",
    number = "1",
    pages = "L5",
    year = "2021"
}

@ARTICLE{valcarce_2025,
       author = {{Valcarce}, Aldo A.~R. and {Catelan}, M{\'a}rcio and {Alves}, S{\^a}nzia and {Gonz{\'a}lez-Bordon}, Felipe},
        title = "{The PGPUC horizontal branch evolutionary tracks}",
      journal = {\aap},
         year = 2025,
        month = jul,
       volume = {699},
          eid = {A368},
        pages = {A368},
          doi = {10.1051/0004-6361/202555099},
archivePrefix = {arXiv},
       eprint = {2506.16562},
 primaryClass = {astro-ph.SR},
       adsurl = {https://ui.adsabs.harvard.edu/abs/2025A&A...699A.368V}
}

@ARTICLE{serenelli_2017,
       author = {{Serenelli}, A. and {Weiss}, A. and {Cassisi}, S. and {Salaris}, M. and
         {Pietrinferni}, A.},
        title = "{The brightness of the red giant branch tip. Theoretical framework, a set of reference models, and predicted observables}",
      journal = {\aap},
         year = 2017,
        month = oct,
       volume = {606},
          eid = {A33},
        pages = {A33},
          doi = {10.1051/0004-6361/201731004},
archivePrefix = {arXiv},
       eprint = {1706.09910},
 primaryClass = {astro-ph.SR},
       adsurl = {https://ui.adsabs.harvard.edu/abs/2017A&A...606A..33S}
}

@BOOK{raffelt_book,
   author = {{Raffelt}, G.~G.},
    title = "{Stars as laboratories for fundamental physics : the astrophysics of neutrinos, axions, and other weakly interacting particles}",
booktitle = {Stars as laboratories for fundamental physics},
     year = 1996,
publisher = {University of Chicago Press, ~QB464.2 .R34 1996},  
   adsurl = {http://adsabs.harvard.edu/abs/1996slfp.book.....R}
}

@ARTICLE{viaux_2013,
   author = {{Viaux}, N. and {Catelan}, M. and {Stetson}, P.~B. and {Raffelt}, G.~G. and 
	{Redondo}, J. and {Valcarce}, A.~A.~R. and {Weiss}, A.},
    title = "{Neutrino and Axion Bounds from the Globular Cluster M5 (NGC 5904)}",
  journal = {\prl},
archivePrefix = "arXiv",
   eprint = {1311.1669},
 primaryClass = "astro-ph.SR",
     year = 2013,
    month = dec,
   volume = 111,
   number = 23,
      eid = {231301},
    pages = {231301},
      doi = {10.1103/PhysRevLett.111.231301},
   adsurl = {http://adsabs.harvard.edu/abs/2013PhRvL.111w1301V}
}

@ARTICLE{straniero_2020,
       author = {{Straniero}, O. and {Pallanca}, C. and {Dalessandro}, E. and {Dom{\'\i}nguez}, I. and {Ferraro}, F.~R. and {Giannotti}, M. and {Mirizzi}, A. and {Piersanti}, L.},
        title = "{The RGB tip of galactic globular clusters and the revision of the axion-electron coupling bound}",
      journal = {\aap},
         year = 2020,
        month = dec,
       volume = {644},
          eid = {A166},
        pages = {A166},
          doi = {10.1051/0004-6361/202038775},
archivePrefix = {arXiv},
       eprint = {2010.03833},
}

@ARTICLE{capozzi_2020,
       author = {{Capozzi}, Francesco and {Raffelt}, Georg},
        title = "{Axion and neutrino bounds improved with new calibrations of the tip of the red-giant branch using geometric distance determinations}",
      journal = {\prd},
         year = 2020,
        month = oct,
       volume = {102},
       number = {8},
          eid = {083007},
        pages = {083007},
          doi = {10.1103/PhysRevD.102.083007},
archivePrefix = {arXiv},
       eprint = {2007.03694},
 primaryClass = {astro-ph.SR},
       adsurl = {https://ui.adsabs.harvard.edu/abs/2020PhRvD.102h3007C}
}

@ARTICLE{carenza_2025PhyR,
       author = {{Carenza}, Pierluca and {Giannotti}, Maurizio and {Isern}, Jordi and {Mirizzi}, Alessandro and {Straniero}, Oscar},
        title = "{Axion astrophysics}",
      journal = {Physics Reports},
         year = 2025,
        month = apr,
       volume = {1117},
        pages = {1-102},
          doi = {10.1016/j.physrep.2025.02.002},
archivePrefix = {arXiv},
       eprint = {2411.02492},
 primaryClass = {hep-ph},
       adsurl = {https://ui.adsabs.harvard.edu/abs/2025PhyR.1117....1C}
}

@INPROCEEDINGS{caputoraffelt_2024,
       author = {{Caputo}, A. and {Raffelt}, G.},
        title = "{Astrophysical Axion Bounds: The 2024 Edition}",
    booktitle = {1st General Meeting and 1st Training School of the COST Action COSMIC WISPers},
         year = 2024,
       volume = {1},
        month = apr,
          eid = {41},
        pages = {41},
       adsurl = {https://ui.adsabs.harvard.edu/abs/2024cacw.confE..41C}
}

@article{Goldsbury:2016ndn,
    author = "Goldsbury, Ryan and Heyl, Jeremy and Richer, Harvey and Kalirai, Jason and Tremblay, Pier-Emmanuel",
    title = "{Constraining white dwarf structure and neutrino physics in 47 Tucanae}",
    eprint = "1602.06286",
    archivePrefix = "arXiv",
    primaryClass = "astro-ph.SR",
    doi = "10.3847/0004-637X/821/1/27",
    journal = "Astrophys. J.",
    volume = "821",
    number = "1",
    pages = "27",
    year = "2016"
}

@article{Isern:2008nt,
    author = "Isern, J. and Garcia-Berro, E. and Torres, S. and Catalan, S.",
    title = "{Axions and the cooling of white dwarf stars}",
    eprint = "0806.2807",
    archivePrefix = "arXiv",
    primaryClass = "astro-ph",
    doi = "10.1086/591042",
    journal = "Astrophys. J. Lett.",
    volume = "682",
    pages = "L109",
    year = "2008"
}

@article{MillerBertolami:2014rka,
    author = "Miller Bertolami, Marcelo M. and Melendez, Brenda E. and Althaus, Leandro G. and Isern, Jordi",
    title = "{Revisiting the axion bounds from the Galactic white dwarf luminosity function}",
    eprint = "1406.7712",
    archivePrefix = "arXiv",
    primaryClass = "hep-ph",
    doi = "10.1088/1475-7516/2014/10/069",
    journal = "JCAP",
    volume = "10",
    pages = "069",
    year = "2014"
}

@article{Isern:2018uce,
    author = "Isern, Jordi and Garcia-Berro, Enrique and Torres, Santiago and Cojocaru, Roxana and Catalan, Silvia",
    title = "{Axions and the luminosity function of white dwarfs: the thin and thick discs, and the halo}",
    eprint = "1805.00135",
    archivePrefix = "arXiv",
    primaryClass = "astro-ph.SR",
    doi = "10.1093/mnras/sty1162",
    journal = "Mon. Not. Roy. Astron. Soc.",
    volume = "478",
    number = "2",
    pages = "2569--2575",
    year = "2018"
}

@article{Fleury:2025ahw,
    author = "Fleury, Leesa and Obertas, Alysa and Richer, Harvey and Heyl, Jeremy",
    title = "{Axion Constraints from White Dwarf Cooling in 47 Tucanae}",
    eprint = "2511.21676",
    archivePrefix = "arXiv",
    primaryClass = "astro-ph.SR",
    month = "11",
    year = "2025"
}

@article{Alberino:2026yxi,
    author = "Alberino, Mart{\'\i}n L. and Miller Bertolami, Marcelo M. and Camisassa, Mar{\'\i}a E. and Caputo, Andrea and Torres, Santiago",
    title = "{New axion bounds derived from the 100-parsec Gaia DR3 white dwarf luminosity function}",
    eprint = "2603.00901",
    archivePrefix = "arXiv",
    primaryClass = "astro-ph.SR",
    month = "3",
    year = "2026"
}

@article{Gontcharov:2026knx,
    author = "Gontcharov, G. A. and Kudashov, A. M. and Ryutina, O. S. and Troitsky, V., S.",
    title = "{Isochrone fitting of Galactic globular clusters - IX. Tip of the red-giant branch for 27 clusters and constraints on new particle physics}",
    eprint = "2608.04801",
    archivePrefix = "arXiv",
    primaryClass = "hep-ph",
    reportNumber = "INR-TH-2026-006",
    month = "8",
    year = "2026"
}

@article{RaffeltDearborn,
  title = {Bounds on hadronic axions from stellar evolution},
  author = {Raffelt, Georg G. and Dearborn, David S. P.},
  journal = {Phys. Rev. D},
  volume = {36},
  issue = {8},
  pages = {2211--2225},
  numpages = {0},
  year = {1987},
  month = {Oct},
  publisher = {American Physical Society},
  doi = {10.1103/PhysRevD.36.2211},
  url = {https://link.aps.org/doi/10.1103/PhysRevD.36.2211}
}

@ARTICLE{Cassisi-Innocenti-Salaris,
       author = {{Cassisi}, Santi and {degl'Innocenti}, Scilla and {Salaris}, Maurizio},
        title = "{The effect of diffusion on the red giant luminosity function `bump'}",
      journal = {\mnras},
         year = 1997,
        month = sep,
       volume = {290},
       number = {3},
        pages = {515-520},
          doi = {10.1093/mnras/290.3.515},
archivePrefix = {arXiv},
       eprint = {astro-ph/9706091},
 primaryClass = {astro-ph},
       adsurl = {https://ui.adsabs.harvard.edu/abs/1997MNRAS.290..515C}
}

@ARTICLE{Cassisi-Salaris,
       author = {{Cassisi}, Santi and {Salaris}, Maurizio},
        title = "{A critical investigation on the discrepancy between the observational and the theoretical red giant luminosity function `bump'}",
      journal = {\mnras},
         year = 1997,
        month = mar,
       volume = {285},
       number = {3},
        pages = {593-603},
          doi = {10.1093/mnras/285.3.593},
archivePrefix = {arXiv},
       eprint = {astro-ph/9702029},
 primaryClass = {astro-ph},
       adsurl = {https://ui.adsabs.harvard.edu/abs/1997MNRAS.285..593C}
}

@ARTICLE{Alongi,
       author = {{Alongi}, M. and {Bertelli}, G. and {Bressan}, A. and {Chiosi}, C.},
        title = "{Effects of envelope overshoot on stellar models.}",
      journal = {\aap},
         year = 1991,
        month = apr,
       volume = {244},
        pages = {95},
       adsurl = {https://ui.adsabs.harvard.edu/abs/1991A&A...244...95A}
}

@article{Cassisi2011,
  author       = {Cassisi, S. and Mar{\'\i}n-Franch, A. and Salaris, M. and Aparicio, A. and Monelli, M. and Pietrinferni, A.},
  title        = {The magnitude difference between the main sequence turn off and the red giant branch bump in Galactic globular clusters},
  journal      = {Astronomy \& Astrophysics},
  volume       = {527},
  pages        = {A59},
  year         = {2011},
  doi          = {10.1051/0004-6361/201016066}
}

@ARTICLE{Riello,
       author = {{Riello}, M. and {Cassisi}, S. and {Piotto}, G. and {Recio-Blanco}, A. and {De Angeli}, F. and {Salaris}, M. and {Pietrinferni}, A. and {Bono}, G. and {Zoccali}, M.},
        title = "{The Red Giant Branch luminosity function bump}",
      journal = {\aap},
         year = 2003,
        month = nov,
       volume = {410},
        pages = {553-563},
          doi = {10.1051/0004-6361:20031272},
archivePrefix = {arXiv},
       eprint = {astro-ph/0308431},
 primaryClass = {astro-ph},
       adsurl = {https://ui.adsabs.harvard.edu/abs/2003A&A...410..553R}
}

@ARTICLE{2010ApJ...712..527D,
       author = {{Di Cecco}, A. and {Bono}, G. and {Stetson}, P.~B. and {Pietrinferni}, A. and {Becucci}, R. and {Cassisi}, S. and {Degl'Innocenti}, S. and {Iannicola}, G. and {Prada Moroni}, P.~G. and {Buonanno}, R. and {Calamida}, A. and {Caputo}, F. and {Castellani}, M. and {Corsi}, C.~E. and {Ferraro}, I. and {Dall'Ora}, M. and {Monelli}, M. and {Nonino}, M. and {Piersimoni}, A.~M. and {Pulone}, L. and {Romaniello}, M. and {Salaris}, M. and {Walker}, A.~R. and {Zoccali}, M.},
        title = "{On the {\ensuremath{\Delta}}V $^{bump}$ $_{HB}$ Parameter in Globular Clusters}",
      journal = {\apj},
         year = 2010,
        month = mar,
       volume = {712},
       number = {1},
        pages = {527-535},
          doi = {10.1088/0004-637X/712/1/527},
archivePrefix = {arXiv},
       eprint = {1002.2074},
 primaryClass = {astro-ph.SR},
       adsurl = {https://ui.adsabs.harvard.edu/abs/2010ApJ...712..527D}
}

@ARTICLE{Joyce_Chaboyer,
       author = {{Joyce}, M. and {Chaboyer}, B.},
        title = "{Investigating the Consistency of Stellar Evolution Models with Globular Cluster Observations via the Red Giant Branch Bump}",
      journal = {\apj},
         year = 2015,
        month = dec,
       volume = {814},
       number = {2},
          eid = {142},
        pages = {142},
          doi = {10.1088/0004-637X/814/2/142},
archivePrefix = {arXiv},
       eprint = {1510.07648},
 primaryClass = {astro-ph.SR},
       adsurl = {https://ui.adsabs.harvard.edu/abs/2015ApJ...814..142J}
}

@article{Meissner2006,
  author       = {Meissner, F. and Weiss, A.},
  title        = {Global fitting of globular cluster age indicators},
  journal      = {Astronomy \& Astrophysics},
  volume       = {456},
  number       = {3},
  pages        = {1085--1096},
  year         = {2006},
  doi          = {10.1051/0004-6361:20065133}
}

@article{Peccei-Quinn1,
  title = {Constraints imposed by $\mathrm{CP}$ conservation in the presence of pseudoparticles},
  author = {Peccei, R. D. and Quinn, Helen R.},
  journal = {Phys. Rev. D},
  volume = {16},
  issue = {6},
  pages = {1791--1797},
  numpages = {0},
  year = {1977},
  month = {Sep},
  publisher = {American Physical Society},
  doi = {10.1103/PhysRevD.16.1791},
  url = {https://link.aps.org/doi/10.1103/PhysRevD.16.1791}
}

@article{Peccei-Quinn2,
  title = {$\mathrm{CP}$ Conservation in the Presence of Pseudoparticles},
  author = {Peccei, R. D. and Quinn, Helen R.},
  journal = {Phys. Rev. Lett.},
  volume = {38},
  issue = {25},
  pages = {1440--1443},
  numpages = {0},
  year = {1977},
  month = {Jun},
  publisher = {American Physical Society},
  doi = {10.1103/PhysRevLett.38.1440},
  url = {https://link.aps.org/doi/10.1103/PhysRevLett.38.1440}
}

@article{Weinberg,
  title = {A New Light Boson?},
  author = {Weinberg, Steven},
  journal = {Phys. Rev. Lett.},
  volume = {40},
  issue = {4},
  pages = {223--226},
  numpages = {0},
  year = {1978},
  month = {Jan},
  publisher = {American Physical Society},
  doi = {10.1103/PhysRevLett.40.223},
  url = {https://link.aps.org/doi/10.1103/PhysRevLett.40.223}
}

@article{Wilczek,
  title = {Problem of Strong $P$ and $T$ Invariance in the Presence of Instantons},
  author = {Wilczek, F.},
  journal = {Phys. Rev. Lett.},
  volume = {40},
  issue = {5},
  pages = {279--282},
  numpages = {0},
  year = {1978},
  month = {Jan},
  publisher = {American Physical Society},
  doi = {10.1103/PhysRevLett.40.279},
  url = {https://link.aps.org/doi/10.1103/PhysRevLett.40.279}
}

@article{Svrcek_2006,
   title={Axions in string theory},
   volume={2006},
   ISSN={1029-8479},
   url={http://dx.doi.org/10.1088/1126-6708/2006/06/051},
   DOI={10.1088/1126-6708/2006/06/051},
   number={06},
   journal={Journal of High Energy Physics},
   publisher={Springer Science and Business Media LLC},
   author={Svrcek, Peter and Witten, Edward},
   year={2006},
   month=June, pages={051–051} }

@article{PRESKILL1983127,
title = {Cosmology of the invisible axion},
journal = {Physics Letters B},
volume = {120},
number = {1},
pages = {127-132},
year = {1983},
issn = {0370-2693},
doi = {https://doi.org/10.1016/0370-2693(83)90637-8},
url = {https://www.sciencedirect.com/science/article/pii/0370269383906378},
author = {John Preskill and Mark B. Wise and Frank Wilczek}
}

@article{ABBOTT1983133,
title = {A cosmological bound on the invisible axion},
journal = {Physics Letters B},
volume = {120},
number = {1},
pages = {133-136},
year = {1983},
issn = {0370-2693},
doi = {https://doi.org/10.1016/0370-2693(83)90638-X},
url = {https://www.sciencedirect.com/science/article/pii/037026938390638X},
author = {L.F. Abbott and P. Sikivie}
}

@article{DINE1983137,
title = {The not-so-harmless axion},
journal = {Physics Letters B},
volume = {120},
number = {1},
pages = {137-141},
year = {1983},
issn = {0370-2693},
doi = {https://doi.org/10.1016/0370-2693(83)90639-1},
url = {https://www.sciencedirect.com/science/article/pii/0370269383906391},
author = {Michael Dine and Willy Fischler}
}

@article{Fields_2020,
   title={Big-Bang Nucleosynthesis after Planck},
   volume={2020},
   ISSN={1475-7516},
   url={http://dx.doi.org/10.1088/1475-7516/2020/03/010},
   DOI={10.1088/1475-7516/2020/03/010},
   number={03},
   journal={Journal of Cosmology and Astroparticle Physics},
   publisher={IOP Publishing},
   author={Fields, Brian D. and Olive, Keith A. and Yeh, Tsung-Han and Young, Charles},
   year={2020},
   month=Mar, pages={010–010} }

@article{XENONnT,
   title={Search for New Physics in Electronic Recoil Data from XENONnT},
   volume={129},
   ISSN={1079-7114},
   url={http://dx.doi.org/10.1103/PhysRevLett.129.161805},
   DOI={10.1103/physrevlett.129.161805},
   number={16},
   journal={Physical Review Letters},
   publisher={American Physical Society (APS)},
   author={Aprile, E. and Abe, K. and Agostini, F. and Ahmed Maouloud, S. and Althueser, L. and Andrieu, B. and Angelino, E. and Angevaare, J. R. and Antochi, V. C. and Antón Martin, D. and Arneodo, F. and Baudis, L. and Baxter, A. L. and Bellagamba, L. and Biondi, R. and Bismark, A. and Brown, A. and Bruenner, S. and Bruno, G. and Budnik, R. and Bui, T. K. and Cai, C. and Capelli, C. and Cardoso, J. M. R. and Cichon, D. and Clark, M. and Colijn, A. P. and Conrad, J. and Cuenca-García, J. J. and Cussonneau, J. P. and D’Andrea, V. and Decowski, M. P. and Di Gangi, P. and Di Pede, S. and Di Giovanni, A. and Di Stefano, R. and Diglio, S. and Eitel, K. and Elykov, A. and Farrell, S. and Ferella, A. D. and Ferrari, C. and Fischer, H. and Fulgione, W. and Gaemers, P. and Gaior, R. and Gallo Rosso, A. and Galloway, M. and Gao, F. and Gardner, R. and Glade-Beucke, R. and Grandi, L. and Grigat, J. and Guida, M. and Hammann, R. and Higuera, A. and Hils, C. and Hoetzsch, L. and Howlett, J. and Iacovacci, M. and Itow, Y. and Jakob, J. and Joerg, F. and Joy, A. and Kato, N. and Kara, M. and Kavrigin, P. and Kazama, S. and Kobayashi, M. and Koltman, G. and Kopec, A. and Kuger, F. and Landsman, H. and Lang, R. F. and Levinson, L. and Li, I. and Li, S. and Liang, S. and Lindemann, S. and Lindner, M. and Liu, K. and Loizeau, J. and Lombardi, F. and Long, J. and Lopes, J. A. M. and Ma, Y. and Macolino, C. and Mahlstedt, J. and Mancuso, A. and Manenti, L. and Marignetti, F. and Marrodán Undagoitia, T. and Martens, K. and Masbou, J. and Masson, D. and Masson, E. and Mastroianni, S. and Messina, M. and Miuchi, K. and Mizukoshi, K. and Molinario, A. and Moriyama, S. and Morå, K. and Mosbacher, Y. and Murra, M. and Müller, J. and Ni, K. and Oberlack, U. and Paetsch, B. and Palacio, J. and Paschos, P. and Peres, R. and Peters, C. and Pienaar, J. and Pierre, M. and Pizzella, V. and Plante, G. and Qi, J. and Qin, J. and Ramírez García, D. and Reichard, S. and Rocchetti, A. and Rupp, N. and Sanchez, L. and dos Santos, J. M. F. and Sarnoff, I. and Sartorelli, G. and Schreiner, J. and Schulte, D. and Schulte, P. and Schulze Eißing, H. and Schumann, M. and Scotto Lavina, L. and Selvi, M. and Semeria, F. and Shagin, P. and Shi, S. and Shockley, E. and Silva, M. and Simgen, H. and Stephen, J. and Takeda, A. and Tan, P.-L. and Terliuk, A. and Thers, D. and Toschi, F. and Trinchero, G. and Tunnell, C. and Tönnies, F. and Valerius, K. and Volta, G. and Wei, Y. and Weinheimer, C. and Weiss, M. and Wenz, D. and Wittweg, C. and Wolf, T. and Xu, D. and Xu, Z. and Yamashita, M. and Yang, L. and Ye, J. and Yuan, L. and Zavattini, G. and Zhong, M. and Zhu, T. and },
   year={2022},
   month=Oct }

@article{Carenza_2021,
   title={Revisiting axion-electron bremsstrahlung emission rates in astrophysical environments},
   volume={103},
   ISSN={2470-0029},
   url={http://dx.doi.org/10.1103/PhysRevD.103.123024},
   DOI={10.1103/physrevd.103.123024},
   number={12},
   journal={Physical Review D},
   publisher={American Physical Society (APS)},
   author={Carenza, Pierluca and Lucente, Giuseppe},
   year={2021},
   month=June }

@article{Straniero_2019,
doi = {10.3847/1538-4357/ab3222},
url = {https://doi.org/10.3847/1538-4357/ab3222},
year = {2019},
month = {aug},
publisher = {The American Astronomical Society},
volume = {881},
number = {2},
pages = {158},
author = {Straniero, Oscar and Dominguez, Inma and Piersanti, Luciano and Giannotti, Maurizio and Mirizzi, Alessandro},
title = {The Initial Mass–Final Luminosity Relation of Type II Supernova Progenitors: Hints of New Physics?},
journal = {The Astrophysical Journal},
}

@article{Raffelt_Weiss,
   title={Red giant bound on the axion-electron coupling reexamined},
   volume={51},
   ISSN={0556-2821},
   url={http://dx.doi.org/10.1103/PhysRevD.51.1495},
   DOI={10.1103/physrevd.51.1495},
   number={4},
   journal={Physical Review D},
   publisher={American Physical Society (APS)},
   author={Raffelt, Georg and Weiss, Achim},
   year={1995},
   month=Feb, pages={1495–1498} }

@article{MARCS,
  author  = {Gustafsson, B. and Edvardsson, B. and Eriksson, K. and
             J{\o}rgensen, U. G. and Nordlund, {\AA}. and Plez, B.},
  title   = {A grid of MARCS model atmospheres for late-type stars - I. Methods and general properties},
  journal = {Astronomy \& Astrophysics},
  year    = {2008},
  volume  = {486},
  number  = {3},
  pages   = {951--970},
  doi     = {10.1051/0004-6361:200809724}
}

@ARTICLE{Thomas_1967,
       author = {{Thomas}, H.-C.},
        title = "{Sternentwicklung VIII. Der Helium-Flash bei einem Stern von 1. 3 Sonnenmassen}",
      journal = {\zap},
         year = 1967,
        month = jan,
       volume = {67},
        pages = {420},
       adsurl = {https://ui.adsabs.harvard.edu/abs/1967ZA.....67..420T}
}

@article{iben_1968,
  author  = {Iben, Icko, Jr.},
  title   = {Age and Initial Helium Abundance of Stars in the Globular Cluster {M15}},
  journal = {Nature},
  year    = {1968},
  volume  = {220},
  pages   = {143--146},
  doi     = {10.1038/220143a0}
}

@ARTICLE{Sarajedini,
       author = {{Sarajedini}, Ata and {Bedin}, Luigi R. and {Chaboyer}, Brian and {Dotter}, Aaron and {Siegel}, Michael and {Anderson}, Jay and {Aparicio}, Antonio and {King}, Ivan and {Majewski}, Steven and {Mar{\'\i}n-Franch}, A. and {Piotto}, Giampaolo and {Reid}, I. Neill and {Rosenberg}, Alfred},
        title = "{The ACS Survey of Galactic Globular Clusters. I. Overview and Clusters without Previous Hubble Space Telescope Photometry}",
      journal = {\aj},
         year = 2007,
        month = apr,
       volume = {133},
       number = {4},
        pages = {1658-1672},
          doi = {10.1086/511979},
archivePrefix = {arXiv},
       eprint = {astro-ph/0612598},
 primaryClass = {astro-ph},
       adsurl = {https://ui.adsabs.harvard.edu/abs/2007AJ....133.1658S}
}

@article{Stetson,
    author = {Stetson, P B and Pancino, E and Zocchi, A and Sanna, N and Monelli, M},
    title = {Homogeneous photometry – VII. Globular clusters in the Gaia era},
    journal = {Monthly Notices of the Royal Astronomical Society},
    volume = {485},
    number = {3},
    pages = {3042-3063},
    year = {2019},
    month = {05},
    issn = {0035-8711},
    doi = {10.1093/mnras/stz585},
    url = {https://doi.org/10.1093/mnras/stz585},
    eprint = {https://academic.oup.com/mnras/article-pdf/485/3/3042/28079473/stz585.pdf},
}

\appendix
\section{Axion production rates}
\label{app:axion_rates}

We summarize here the numerical recipes for the computation of the rates of axion production mechanisms relevant to this work, namely Compton and Bremsstrahlung. We also include the Primakoff rate. For a more detailed discussion see \cite{Straniero_2019} and references hereafter. For a general review on the topic, see \citep{raffelt_book}\\
\subsection{Some general definitions}

The density of the j ions is given by:
\begin{equation}
    n_j=\frac{\rho}{m_u}\frac{X_j}{A_j}
\end{equation}
similarly the electron density is:
\begin{equation}
    n_e=\frac{\rho}{m_u\mu_e}
\end{equation}
where $1/\mu_e=\sum X_jZ_j/A_j$ and $m_u=1.66\times10^{-24}$g
\subsubsection{Primakoff}
  The Primakoff mechanism is the conversion of photons into axions in the presence of a magnetic field or interaction with charged particles such as electrons and ions  \citep{RaffeltDearborn}. It is controlled by the axion-photon coupling. The rate is calculated by adding the contribution from the interaction with ions and electrons. The ions are assumed to be non-degenerate (true for all the stellar stages of interest) while the electrons can be non-degenerate (MS) or degenerate (RGB). We take into account the stellar plasma screening by "renormalizing" the propagator by the structure function: 
\begin{equation}
    \frac{1}{q^4}\longrightarrow \frac{1}{q^4}S(q^2)= \frac{1}{q^2(q^2+\kappa_{\text{D/TF}}^2)}
\end{equation}
where 
\begin{equation}
    \kappa_{\text{D}}^2=\frac{4\pi Z^2\alpha n}{T}
\end{equation}
is the Debye-Hünckel screening-length and 
\begin{equation}
    \kappa_{\text{TF}}^2=\frac{4 Z^2\alpha m p_F}{\pi }
\end{equation}
is the Thomas-Fermi screening length for a fully degenerate medium.\\
To account for intermediate degeneracy we set the degeneracy parameter such that:
\begin{equation}
    \kappa^2=\theta_{\text{deg}}\kappa_{\text{D}}^2
\end{equation}
We fit this degeneracy parameter as a function of the temperature and density of the electron gas. Introducing the dimensionless parameter
\begin{equation}
    z=\frac{\rho}{T^{3/2}\mu_e}
\end{equation}

\begin{equation}
{\small
\theta_{\rm deg} =
\left\{
\begin{array}{l}
1\hspace{2.9cm}\text{for }z\leq5.45\times10^{-11}\\[1pt]
0.63+0.3\tan^{-1}\!\left(
0.65-9316z^{0.48}
+\dfrac{0.019}{z^{0.212}}
\right)\\[-1pt]
\hspace{3.cm}\text{for }5.45\times10^{-11}<z\leq7.2\times10^{-8}\\[1pt]
4.78\times10^{-6}z^{-0.667}
\hspace{0.4cm}\text{for }z>7.2\times10^{-8}
\end{array}
\right.
}
\raisetag{15pt}
\end{equation}

Writing:
\begin{align}
    &y_{\text{pl}}=\frac{\omega_{\text{pl}}}{T}=\frac{(\rho/\mu_e)^{1/2}}{(1+(1.02\times10^{-6}\rho/\mu_e)^{2/3})^{1/4}} \\
    &y_{\text{ions}}=2.57\times 10^{10}\left(\frac{\rho}{T^3}\sum_{\text{ions}}\frac{Z_j^2X_j}{A_j}\right)^{1/2}\\
    &y_{\text{el}}=2.57\times 10^{10}\theta_{\text{deg}}\left(\frac{\rho}{T^3}\sum_{\text{ions}}\frac{Z_jX_j}{A_j}\right)^{1/2}\\
    &y_s=(y_{\text{ions}}^2+y_{\text{el}}^2)^{1/2} 
\end{align}.
\begin{align}
   & f(y_0,y_1)=\frac{1}{4\pi}\int_{y_0}^{\infty}dy\frac{y^2(y^2-y_0^2)^{1/2}}{e^y-1}I(y,y_0,y_1)  \label{original integral}\\
    &I = \int_{-1}^{1} \frac{1 - x^2}{(r - x)(r + s - x)} \, dx \label{I}\\
    &r = \frac{2y^2 - y_0^2}{2y\sqrt{y^2 - y_0^2}} \\
    &s = \frac{y_1^2}{2y\sqrt{y^2 - y_0^2}}
\end{align}.

The energy loss rate due to Primakoff axion production is given by:
\begin{equation}
    \epsilon = 4.71\times10^{-31}g_{10}^2T^4\left[\sum_{\text{ions}}(Z_j^2+\theta_{\text{deg}}Z_j)\frac{X_j}{A_j}\right]f(y_{\text{pl}},y_s)\frac{\text{erg}}{\text{g.s}}
\end{equation}

\subsubsection{Compton}
Compton axion production refers to axion production via the scattering of thermal photons off electrons.  \citep{Raffelt_Weiss}. It is controlled by the coupling with electrons.
The rate is given by:
\begin{equation}
    \epsilon=\theta_{\text{deg}}\frac{n_e}{\rho}\int2\frac{\mathbf{d^3k}}{(2\pi)^3}\frac{\sigma\omega}{e^{\omega/T}-1}
\end{equation}
where $\sigma=\frac{1}{3}\alpha(g_{ae}/m_e)^2(\omega/m_e)^2$ is the Compton cross section, and $\theta_{\text{deg}}$ is the degeneracy parameter described above. Numerically, the rate becomes
\begin{equation}
    \epsilon=\theta_{\text{deg}}2.66\times10^{-48}(g_{ae}/10^{-13}\text{eV})^2\frac{T^6}{\mu_e}
\end{equation}

\subsubsection{Bremsstrahlung}
Analogously to classical Bremsstrahlung, it is the axion emission by a charged particle interacting in an electromagnetic field. It is controlled by the coupling with electrons. We follow \citep{Raffelt_Weiss}.
The non degenerate limit reads:
\begin{equation}
        \epsilon_{\text{nd}}=4.7\times10^{-25}g_{13}^2T^{2.5}\frac{\rho}{\mu_e}\sum\frac{X_j}{A_j}\left(Z_j^2+\frac{Z_j}{\sqrt{2}}\right)\text{erg.g$^{-1}$.s$^{-1}$}
\end{equation}
The fully degenerate limit is given by:
\begin{equation}
   \epsilon_{\text{d}}=8.6F\times10^{-33}g_{13}^2T^4\left(\sum\frac{X_jZ^2_j}{A_j}\right)
\end{equation}
\begin{equation}
   F=\frac{2}{3}\log\left(\frac{2+\kappa^2}{\kappa^2}\right)+\left[\frac{2+5\kappa^2}{15}\log\left(\frac{2+\kappa^2}{\kappa^2}\right)-\frac{2}{3}\right]\beta_F^2
\end{equation}
\begin{equation}
   \kappa^2=k_D^2/2p_F^2
\end{equation}
where $\beta_F\equiv p_F/E_F$ and $p_F$ and $E_F$ are, respectively, the Fermi momentum and Fermi level.\\

To account for partial degeneracy, we use the interpolation formula from \citep{Raffelt_Weiss}
\begin{equation}
    \epsilon=(\frac{1}{\epsilon_{\text{ND}}}+\frac{1}{\epsilon_{\text{D}}})^{-1}
\end{equation}
It has to be noted that the complete computation has been reevaluated in \cite{Carenza_2021}. The authors found that in the range of conditions relevant for RGB stars, the interpolation formula above underestimates the axion luminosity by roughly 10\%, thus leading to a slightly overly pessimistic bound.

\end{document}